\documentclass[aps,pra,twocolumn,nofootinbib,floatfix,longbibliography,superscriptaddress]{revtex4-2}

\usepackage{amsmath,amssymb,amsthm,bbm,bm,color,dsfont,float,graphicx,hyperref,makecell,mathrsfs,mathtools,nicefrac,pgfplots,physics,tikz,times,txfonts}
\usepackage{comment}
\usepackage[normalem]{ulem}  

\definecolor{equationcolor}{RGB}{222,94,100}
\definecolor{alecolor}{RGB}{238,33,80}

\hypersetup{urlcolor=equationcolor, colorlinks=true,citecolor=equationcolor,linkcolor=equationcolor} 
\pgfplotsset{compat=1.18} 

\DeclareFontFamily{U}{mathb}{\hyphenchar\font45}
\DeclareFontShape{U}{mathb}{m}{n}{
	<-6> mathb5 <6-7> mathb6 <7-8> mathb7
	<8-9> mathb8 <9-10> mathb9
	<10-12> mathb10 <12-> mathb12
}{}
\DeclareSymbolFont{mathb}{U}{mathb}{m}{n}
\DeclareMathSymbol{\ggcurly}{\mathrel}{mathb}{"CF}

\renewcommand{\v}[1]{\ensuremath{\boldsymbol #1}}

\makeatletter
\def\blfootnote{\gdef\@thefnmark{}\@footnotetext}
\makeatother

\theoremstyle{plain}
\newtheorem{thm}{Theorem}

\newtheorem{prop}{Proposition}

\def\>{\rangle}
\def\<{\langle}

\def\cblue{\color{black}}

\def\blk{\color{black}}
\def\cblu{\color{black}}

\newlength\myindent
\begin{document}
\newcommand{\cologne}{Department Mathematik/Informatik--Abteilung Informatik,\protect\\ Universit\"at zu K\"oln, Albertus-Magnus-Platz, 50923 K\"oln, Germany}
\newcommand{\lanl}{Theoretical Division (T-4), Los Alamos National Laboratory, Los Alamos, New Mexico 87545, USA.}
\newcommand{\iiserk}{Department of Physical Sciences, Indian Institute of Science Education
and Research Kolkata,\\ Mohanpur, West Bengal 741246, India}
\author{Tanmoy Biswas}
\email{tanmoy.biswas23@gmail.com}
\affiliation{\cologne}
\affiliation{\lanl}
\author{Chandan Datta}
\affiliation{\iiserk}
\author{Luis Pedro Garc\'ia-Pintos}
\affiliation{\lanl}

\title{All steerable quantum correlations can provide thermodynamic advantages in cooling}

\date{\today}
\begin{abstract}
     The removal of heat generated during computation poses a major challenge for both classical and quantum computation and information processing. In particular, removal of this heat is intrinsically tied to one of the fundamental requirements of quantum computation—the need to reset the system to a pure state before computation. Therefore, efficient cooling is of paramount importance, not only for deepening our understanding of thermodynamics in the quantum regime but also for advancing the development of modern quantum technologies. In this letter, we devise a cooling task that exploits steerability—a fundamental feature of quantum correlations—to demonstrate a provable quantum thermodynamic advantage over classically correlated scenarios where steerability is absent. We quantify the advantage by the ratio of the amount of heat removed using steerable (quantum) correlations to that obtained with unsteerable (classical) correlations. Specifically, we show that steerable quantum correlations always yield a thermodynamic advantage in a cooling task over their classical counterparts. Remarkably, we further establish that the maximum achievable advantage is directly related to a geometric measure of steerability called \emph{steerability robustness}. Our results suggest that this thermodynamic advantage can be interpreted as a witness of steerability. Finally, we present examples in which the advantage grows with the dimension of the underlying system. 
\end{abstract}
\maketitle

\textit{Introduction.---}Cooling quantum systems is a fundamental thermodynamic task, essential both for practical applications in modern quantum technologies and for foundational investigations \cite{divincenzo2000physical,Deutsch_PRXQ,Aspect2023,KnillZurek1998,Preskill1998Reliable,schulman1998scalablenmrquantumcomputation,SchulmanVazirani, OliveiraJr2025,sarkar2025}. \cblue Computation—whether classical or quantum—can be fundamentally regarded as a thermodynamic process that converts free energy into waste heat and logical work, as established in the seminal works of Bennett~\cite{Bennett_1982} and Landauer~\cite{Landauers_principle2,Landauers_principle}. \blk
 In particular, in quantum computation, qubits operate at the scale where even weak thermal noise can degrade superposition and entanglement, which are the essential resources for information processing. Beyond this fundamental limitation, additional heat is generated due to experimental constraints and the characteristics of the specific physical platform used to implement the computation \cite{Exp1PRXQ,Pop2014,Bruzewicz2019,Slussarenko2019}. Therefore, a quantum processor performing computation must be maintained at very low temperatures by continuously removing the generated heat. Efficient heat removal thus remains a key challenge for the realization of scalable quantum computers.

  Inspired by the operation of a microscopic Otto engine \cite{Equivelence_Uzdin_Kosloff,Kosloff_Rezek2017,Zhang2014PRL,ZhangPRA,BiswasPRL,BiswasPRE,Biswas2022extractionof,Lobejko_Biswas2026} and refrigerator~\cite{Kosloff_Feldman2000,Kosloff_Feldman2010,Abah_2016,AbahPRR}, we propose a cooling protocol fueled by resources arising from quantum correlations shared between two spatially separated parties, enabling the withdrawal of substantially more heat from a thermal bath than is possible using resources based solely on classical correlations. We show that steerability is the key resource underlying these advantages, quantified by the ratio of heat removed with and without steerability.

The concept of steerability was first introduced by Schrodinger in response to the Einstein-Podolsky-Rosen (EPR) paradox~\cite{EPR_paradox,Schrodinger_1935,Schrodinger_1936}. Steerability of quantum correlations, also known as \emph{quantum steering}, is a feature of quantum correlations whereby a spatially separated observer can influence the set of conditional quantum states, referred to as \emph{assemblages}, accessible to a distant party as a result of performing uncharacterized measurements, in a manner that cannot be explained by local causal models.~\cite{Wiseman_steering, SteeringRPP, SteeringRevModPhys}. Here, \emph{uncharacterized measurements} refer to measurements whose specific measurement operators are unknown to the party, so that the corresponding conditional states are not accessible. When both parties perform uncharacterized measurements, the scenario corresponds to the device-independent framework of Bell nonlocality~\cite{Brunner_nonlocality_review, ScaraniBook, ScaraniOutlook, Acin2016, MassarPironio}. Quantum correlations in entangled states that violate Bell inequalities are always steerable, but the converse does not necessarily hold.

Quantum steering encapsulates intrinsically quantum features that have no classical counterpart. Previous studies have demonstrated that steerability of quantum correlations offers significant advantages across a wide range of information-theoretic tasks, including quantum key distribution~\cite{Branciard_QKD,Ma2012}, quantum key authentication~\cite{Mondal_QKA}, randomness certification~\cite{Law_2014,Passaro_2015}, sub-channel discrimination~\cite{Piani2009,Piani_subchannel}, distinguishing quantum measurements~\cite{Datta_2021}, characterization of measurement incompatibility~\cite{Chen_incompatibility,Cavalcanti_incompatibility}, secret sharing~\cite{Xiang_secret_sharing,Kogias2015}, quantum metrology~\cite{Yadin_metrology}, quantum teleportation~\cite{Reid2013,He2015,Pirandola2015,Cavalcanti_teleportation} and work extraction~\cite{Rio2011,Steering_Engines,Steering_Engines_expt,BiswasDattaPRL,Hsieh_PRL,hsieh2024generalquantumresourcesprovide,hsieh2025completecharacterisationstateconversions}.

 
  \cblue \blk
In this Letter, we show that for every steerable assemblage there exists a cooling task where steerability acts as a genuine resource, enabling greater heat withdrawal from the bath than any unsteerable scenario. Moreover, the maximum achievable advantage is lower bounded by the steerability robustness—a geometric quantifier—thereby allowing this advantage itself to serve as a witness of steerability.

In previous studies, robustness has been extensively explored across various resource theories—including entanglement \cite{Robustness_Vidal,Steiner2003,Brandao2005}, coherence \cite{Napoli2016,Biswas2017Visibility}, asymmetry \cite{AsymmetryPiani}, non-stabilizerness \cite{Howard,Winter2022}, and measurement informativeness \cite{Resorcetheory_Informativeness}—and linked to operational tasks such as state and subchannel discrimination \cite{Oszmaniec2019operational,TakagiPRL,TakagiPRX,Haapasalo_2015,Chitambar_review,gour2024resourcesquantumworld,Salazar2021optimalallocationof}. To the best of our knowledge, our work provides the first operational interpretation of robustness in a thermodynamic setting. As a concrete example, motivated by the connection between steerability and measurement incompatibility~\cite{Uolaonetoone2015}, we consider assemblages generated by mutually unbiased bases (MUBs) measured on one subsystem of a maximally entangled state~\cite{Ivonovic_1981,WOOTTERS1989363}. These MUBs achieve maximal incompatibility and thus maximal steerability robustness, which scales as $\mathcal{O}(\sqrt{d})$ with the Hilbert space dimension $d$. Consequently, the resulting quantum cooling advantage grows with system size.


\begin{figure}[t]
    \centering
    \includegraphics[width=4cm]{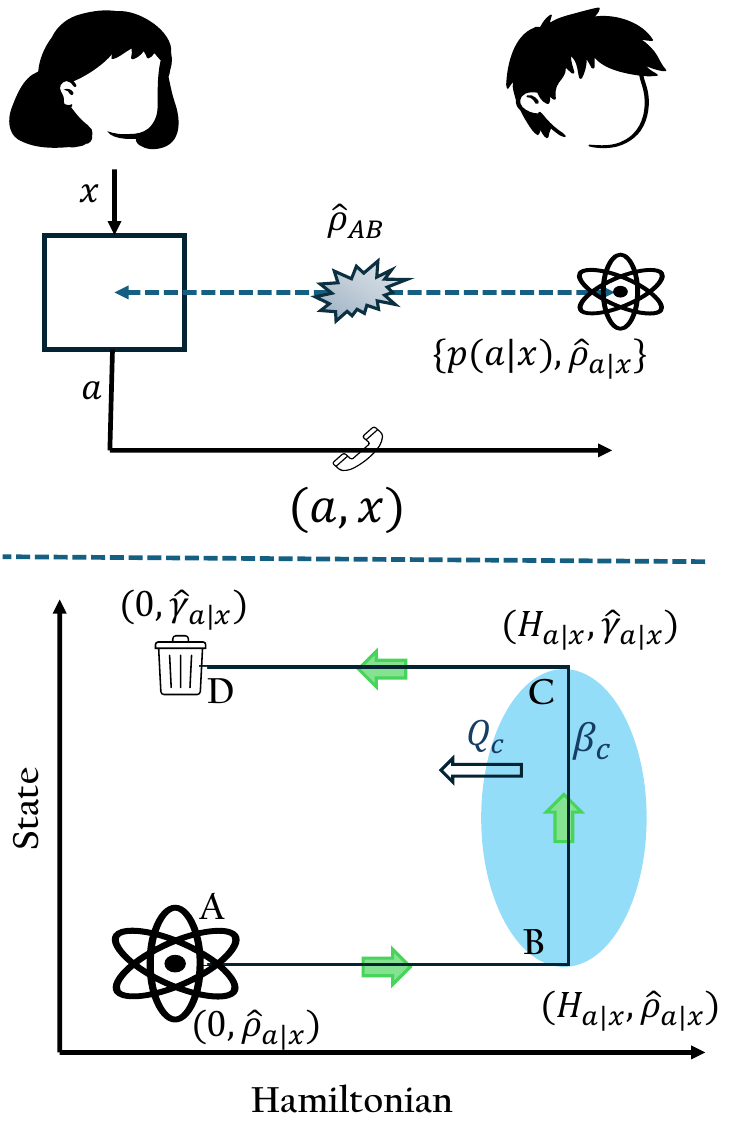}
    \caption{\label{fig:Steering_task} 
In the upper panel, Alice and Bob share a bipartite state $\hat{\rho}_{AB}$. Alice samples $x\in\{0,\ldots,n-1\}$ uniformly, performs measurement $M_x$, and communicates $(a,x)$ to Bob, preparing the conditional state $\hat{\rho}_{a|x}$ with probability $p(a|x)$ (see Eqs.~\eqref{conditional_state} and \eqref{conditional_prob}), which serves as a resource or fuel for heat withdrawal from the bath. In the lower panel, upon receiving $(a,x)$, Bob quenches his Hamiltonian from $H=0$ to $H_{a|x}$ followed by thermalization with a bath at inverse temperature $\beta$ to extract heat (step (2)), and quenches back to $H=0$ (step (3)). The average amount of the withdrawal heat is
$Q_c = \frac{1}{n} \sum_{a,x} p(a|x)\left[\Tr(\hat{\gamma}_{a|x} H_{a|x}) - \Tr(\hat{\rho}_{a|x} H_{a|x})\right]$.}
\end{figure}

\textit{Main results.---}Cooling task using classical and quantum resources: We consider a cooling task that aims to withdraw heat from a thermal bath, similar to an Otto refrigerator, which transfers heat from a cold to a hot bath using external work. In contrast, we focus on a single bath and investigate heat withdrawal fueled by resources arise from classical and quantum correlations, highlighting potential quantum advantages in cooling.

Assume that Bob has access to a heat bath at inverse temperature \(\beta\), from which he seeks to withdraw heat with the assistance of Alice, who is spatially separated. Alice and Bob share a joint quantum state \(\hat{\rho}_{AB}\). We consider Bob's system to have finite dimension \(d < \infty\), with its initial Hamiltonian specified as \(H = 0\).

\begin{figure}[t]
    \centering
    \includegraphics[width=4.5cm]{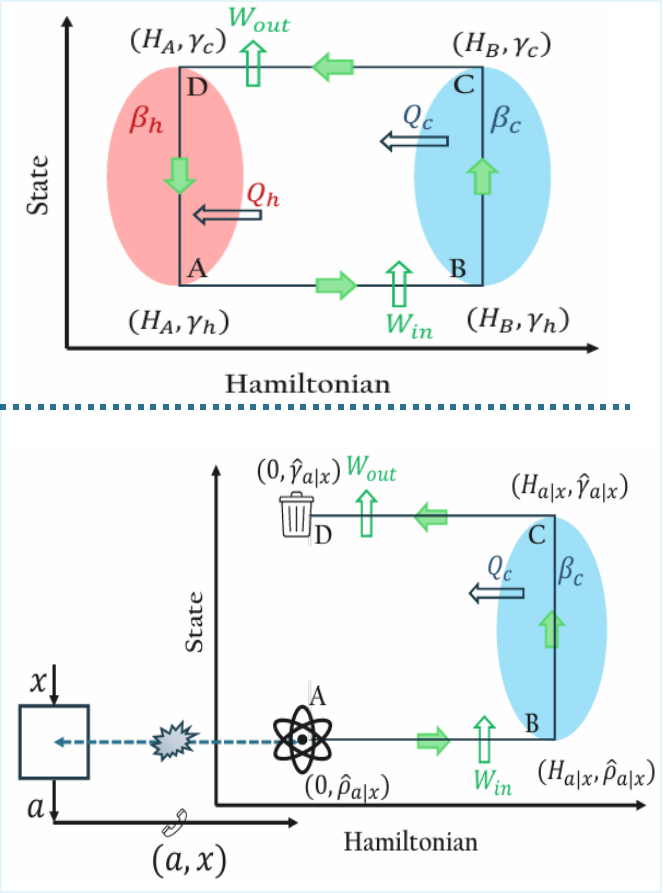}
    \caption{\label{fig:Otto_vs_correlation_driven} In the upper panel, we depict the Otto refrigeration cycle in the space of system state and Hamiltonian, where heat $Q_c$ is extracted from the cold bath and $Q_h$ is released to the hot bath, requiring total work $Q_h + Q_c = W_{\mathrm{in}} + W_{\mathrm{out}}$, with $W_{\mathrm{in}} = \Tr(H_B \gamma_h) - \Tr(H_A \gamma_h)$ and $W_{\mathrm{out}} = \Tr(H_A \gamma_c) - \Tr(H_B \gamma_c)$. In the lower panel, we show the correlation-assisted cooling protocol, where assemblages from steerable correlations enable greater heat withdrawal $Q_c$ than those from unsteerable correlations, demonstrating a quantum advantage.    
}
\end{figure}
To prepare resources that Bob can use as fuel for cooling, Alice samples an input \(x\) from the set \(\mathcal{X} = \{0, \ldots, n-1\}\) and performs the measurement \(M_x\) on the bipartite state \(\hat{\rho}_{AB}\). Each measurement \(M_x\) is described by a positive operator-valued measure (POVM) \(M_x = \{ M_{a|x} \}_{a=0}^{o-1}\), where $o$ is the total number of outcomes.  
Upon obtaining the outcome \(a\), she communicates the pair \((a, x)\) to Bob via a classical communication channel.  
As a result of Alice's measurement, Bob’s subsystem collapses to the conditional state  
\begin{equation}\label{conditional_state}
    \hat{\rho}_{a|x}= \Tr_{A}\!\left[\left(M_{a|x}\otimes\mathbb{I}\right)\hat\rho_{AB}\right]/p(a|x),
\end{equation}
with the corresponding probability  
\begin{equation}\label{conditional_prob}
    p(a|x)= \Tr\!\left[\left(M_{a|x}\otimes\mathbb{I}\right)\hat\rho_{AB}\right].
\end{equation}

To withdraw heat from the bath, Bob carries out the following procedure:

\begin{itemize}
    \item [1)] After receiving $a$ and $x$, Bob quenches his initial Hamiltonian \(H = 0\) to \(H_{a|x}\).
    \item [2)] Next, Bob thermalizes his system \(\hat{\rho}_{a|x}\) \cblue irreversibly, \blk to the thermal state \(\hat{\gamma}_{a|x}\) with respect to the Hamiltonian \(H_{a|x}\) using a heat bath at finite non-zero temperature \(T\) i.e., $0<T<\infty$. \cblue This process results in the exchange of heat from the bath, which is given by the difference between the average energies of the final and initial states associated with this step, i.e.,
\begin{align}\label{Eq_Heat_withdrawn}
     \quad \quad Q_c(\hat{\rho}_{a|x}, H_{a|x}) 
     &=  \Tr \bigl( H_{a|x}\hat{\gamma}_{a|x} \bigr)
     - \Tr \bigl( H_{a|x}\hat{\rho}_{a|x} \bigr),\\
\text{where}\quad
\hat{\gamma}_{a|x} 
&:= \frac{e^{-\beta H_{a|x}}}{\Tr\left(e^{-\beta H_{a|x}}\right)}, 
\quad \text{with}\quad 
\beta=(k_B T)^{-1},
\end{align}
and $k_B$ denotes the Boltzmann constant. \blk
    \item [3)] In the final step, Bob restores the system to its initial Hamiltonian \(H = 0\) through another quench performed without any interaction with the heat bath.
\end{itemize}

Bob’s protocol  follows a four-stroke Otto refrigerator cycle to extract heat from the bath. Unlike the standard cyclic operation that requires re-initialization via an additional bath, we do not enforce a return to the initial state; instead, the protocol is restarted by Alice performing measurements on $\hat{\rho}_{AB}$, leveraging shared correlations.

\cblue To contrast classical and quantum scenarios, we assume that Alice and Bob share many copies of a joint state $\rho_{AB}$, from which the resources for heat withdrawal are generated. Since our aim is to highlight the advantage arising from quantumness, we do not account for the preparation/re-initialization cost of these resources (i.e., preparing $\rho_{AB}$). This follows the standard approach in studies of quantum advantage, where the focus is on how the presence of a quantum resource enhances the performance of a given task over classical scenario. 

\cblue Considerable effort has been devoted to identifying thermodynamic signatures of quantum correlations, particularly in the contexts of work extraction, work storage, and anomalous heat flow. In work extraction and storage protocols, correlations between subsystems can enhance performance by enabling greater work extraction or storage when global operations on the correlated systems are allowed \cite{GRSprl2022,FustratingQB,LlobetPRX,Manzano2018,Campisi2019}. Another striking manifestation is anomalous heat flow, wherein correlations permit heat to flow from a colder body to a hotter one, in apparent violation of the standard thermodynamic expectation, typically captured through a modified form of the \emph{Clausius inequality}~ \cite{Partovi,JenningsRudolphAHF,Micadei2019,PRXQComar}. In contrast, our framework departs fundamentally from these settings. We do not consider direct work extraction or heat exchange between correlated subsystems (i.e., Alice's and Bob's systems). Instead, we focus on correlations shared between Alice and Bob as a ingredient that results out-of-equilibrium resources or “fuel,”  which Bob can subsequently exploit to withdraw heat from a thermal bath. Importantly, in our scenario, Bob’s system remains uncorrelated with the bath from which heat flow is initiated, throughout the entire process. 
\blk

\emph{Heat and work associated with the cooling protocol:} The amount of heat exchanged with the heat bath, averaged over all possible inputs \(x\) and
outcomes \(a\), is given by  
\begin{align}\label{Eq_Avg_heat_withdrawn}
    &Q_c\left(\{\rho_{a|x}\},\{H_{a|x}\}\right) 
   = \frac{1}{n}\sum_{a,x} p(a|x)\left[\Tr \bigl( H_{a|x}\hat{\gamma}_{a|x} \bigr)-\Tr \bigl( H_{a|x}\hat{\rho}_{a|x} \bigr)\right] \nonumber\\
    =& \frac{1}{n}\sum_{a,x} \left[ \Tr \bigl( H_{a|x}\gamma_{a|x} \bigr)-\Tr \bigl( H_{a|x}\rho_{a|x} \bigr)\right],
\end{align}
where we define
\begin{equation}\label{Eq_assemblages}
    \rho_{a|x} := p(a|x)\hat{\rho}_{a|x},
    \qquad
    \gamma_{a|x} := p(a|x)\hat{\gamma}_{a|x}.
\end{equation}
\cblue If $Q_c(\{\hat{\rho}_{a|x}\}, \{H_{a|x}\}) \geq 0$, heat is withdrawn from the bath; whereas if $Q_c(\hat{\rho}_{a|x}, H_{a|x}) < 0$, heat is released into the bath.  Since our primary interest lies in the withdrawal of heat from the bath, we restrict our attention to the former scenario. \blk The prefactor \(1/n\) in Eq.~\eqref{Eq_Avg_heat_withdrawn} arises from the uniform sampling of \(x\) over the set \(\mathcal{X} = \{0, \ldots, n-1\}\). Observe that for all $x$, $\sum_{a}\rho_{a|x}=\mathrm{Tr}_{A}(\hat{\rho}_{AB})$; we refer to any such collection of sub-normalized states $\{\rho_{a|x}\}_{a,x}$ as an \emph{assemblage}~\cite{SteeringRPP}. Their traces yield the probabilities of obtaining outcome \(a\) for the measurement setting \(x\) sampled by Alice, i.e., \(\Tr(\rho_{a|x}) = \Tr(\gamma_{a|x}) = p(a|x)\). Consequently, an assemblage \(\{\rho_{a|x}\}_{a,x}\) can be interpreted as a source, indexed by \(x\), producing the normalized quantum state \(\hat{\rho}_{a|x}\) with probability \(\Tr(\rho_{a|x})\). 

\cblue On the other hand, the average amount of work performed on the system during the first and third steps can be obtained by summing the differences between the average energies of the final and initial states associated with these two steps respectively. This gives
\begin{align}
W\left(\{\rho_{a|x}\},\{H_{a|x}\}\right)
=& \frac{1}{n}\sum_{a,x}
\Big[
\Tr(H_{a|x}\rho_{a|x})
-
\Tr(H_{a|x}\gamma_{a|x})
\Big].
\end{align}

When heat is withdrawn from the bath, i.e., $Q_c(\{\rho_{a|x}\},\{H_{a|x}\})>0$, the amount of work performed on the system is equal in magnitude but opposite in sign to the withdrawn heat, namely $W_c(\{\rho_{a|x}\},\{H_{a|x}\})=-Q_c(\{\rho_{a|x}\},\{H_{a|x}\})$. \cblu This implies that, while extracting heat from the bath, the protocol simultaneously extracts work. At first sight, this may appear to contradict the operation of a standard Otto refrigerator, where external work must be supplied to remove heat from the cold bath. 

The key distinction lies both in the nature of the resource that drives the process and in the protocol itself. As discussed above, our protocol is powered by an out-of-equilibrium resource, which serves as the fuel enabling heat withdrawal from the bath. Moreover, the heat-withdrawal process considered here is not cyclic with respect to the fuel, as the protocol does not return the fuel to its initial state. By contrast, an Otto refrigerator operates as a closed thermodynamic cycle between two equilibrium reservoirs at different temperatures and consumes external work to transfer heat from the cold reservoir to the hot one. Since the proposed protocol does not return the fuel to its initial state, we do not account for the work required to restore the fuel, unlike in an Otto refrigerator. The objective of this Letter is to show that quantum resources used as fuel (with the notion of quantumness defined precisely in the following paragraph) can outperform classical resources in the task of withdrawing heat from a thermal bath.\blk 

Since we take the initial Hamiltonian of Bob's system, which fuels the process, to be zero, the heat withdrawn is equal in magnitude and opposite in sign to the work associated with the protocol, irrespective of the nature of the resources involved.

\emph{Quantum thermodynamic advantage in cooling and steerability robustness:} To identify signatures of \emph{quantumness}, we introduce the notion of steerability.  
If the postmeasurement assemblages $\{\rho_{a|x}\}_{a,x}$ can be written as  
\begin{equation}\label{Eq_LHS_decomposition}
    \rho_{a|x} = \int d\lambda\, p(\lambda)\, p(a|x,\lambda)\, \sigma_\lambda
    \qquad \forall\, a,x,
\end{equation}
then $\{\rho_{a|x}\}_{a,x}$ is called \emph{unsteerable} assemblage.  This decomposition in Eq. \eqref{Eq_LHS_decomposition} is referred to as the \emph{local hidden state} (LHS) decomposition. An assemblage that is not unsteerable is \emph{steerable}. We denote the set of all assemblages by $\mathcal{L}$.

Unsteerable assemblages are regarded as \emph{classical}, as they admit a local hidden state (LHS) decomposition, meaning that all measurement statistics can be reproduced using classical correlations (e.g., those arising from separable states \footnote{\cblue In this article, we take classical correlations as those that give rise only to unsteerable assemblages upon performing any local measurement by Alice. Since correlations arising from any separable state lead to unsteerable assemblages for all local measurements, we treat such correlations as classical. However, we note that in the literature, classically correlated states are sometimes more narrowly defined as separable states with zero discord, i.e., states of the form ( $\sum_{i} p_i \ketbra{i} \otimes \rho_i$ ).
\cite{Adesso_2016}}\blk). In contrast, assemblages \(\{\rho_{a|x}\}_{a,x}\) that do not admit such a decomposition (see Eq.~\eqref{Eq_LHS_decomposition}) can not be generated by local measurements on separable states and thus require a shared entangled state \(\hat{\rho}_{AB}\).

In this work, we compare the heat extracted from a bath using unsteerable (classical) and steerable (quantum) assemblages. For a fair comparison, we impose that both scenarios share the same probability distribution:
\begin{equation}\label{Eq_Fixed_weight_eqvt}
    \forall a,x \quad \Tr(\rho_{a|x}) = \Tr(\sigma_{a|x})=p(a|x).
\end{equation}
Here, \(\sigma_{a|x}\) denotes an unsteerable assemblage admitting an LHS decomposition (Eq.~\eqref{Eq_LHS_decomposition}), while \(\rho_{a|x}\) is steerable \cite{Hsieh_PRL}. Operationally, Bob accesses quantum resources (normalized states \(\hat{\rho}_{a|x}\)) with probability \(p(a|x)\) to extract heat, and we compare this with the classical case where he instead uses non-steerable states \(\hat{\sigma}_{a|x}\) with the same probabilities. For a given \(\rho_{a|x}\), we define the set of compatible unsteerable assemblages as
\begin{equation}\label{defn_Lrho}
    \mathcal{L}_{\rho}:=
    \Big\{
    \{\sigma_{a|x}\}_{a,x}\in\mathcal{L}
    \;\big|\; \Tr(\rho_{a|x})=\Tr(\sigma_{a|x})=p(a|x)
    \quad\forall \;a,x\Big\}.
\end{equation}

Cooling task using classical and quantum resources: To quantify the degree of steerability of states, we consider the 
\emph{steering robustness of an assemblage}~\cite{Piani2009,Skrzypczyk2014}.
Analogous to entanglement robustness, the steering robustness captures the minimal amount of noise that must be mixed with a given assemblage to make it unsteerable. The definition of the steerability robustness of an assemblage is 
\begin{align}\label{Eq:Robustness_of_steering}
    \mathcal{R}(\{\rho_{a|x}\}_{a,x}) &= \min_s\Bigg\{\left\{\frac{\rho_{a|x}+s\tau_{a|x}}{1+s}\right\}_{a,x}\in \mathcal{L}, \nonumber\\&\text{where $\{\tau_{a|x}\}_{a,x}$ is an arbitrary assemblage}\Bigg\}.
\end{align}
Here, $\mathcal{L}$ denotes the set of all unsteerable assemblages. It straightforwardly follows that $\mathcal{R}(\{\rho_{a|x}\}_{a,x})=0$ if and only if $ \{\rho_{a|x}\}_{a,x}\in\mathcal{L}$, and it is strictly greater than $0$, otherwise.

We present the main result of this Letter, establishing a thermodynamic advantage in cooling with steerable over unsteerable assemblages satisfying Eq.~\eqref{Eq_Fixed_weight_eqvt}. We quantify this advantage as the ratio of heat removal using a steerable assemblage to the optimal unsteerable one, and relate it to the steering robustness in Eq.~\eqref{Eq:Robustness_of_steering}. 
\cblue
\begin{thm}\label{Robustness_assemblage}
For any steerable assemblage $\{\rho_{a|x}\}_{a,x} \notin \mathcal{L}$,
\begin{align}
    \xi_{\max}\bigl(\{\rho_{a|x}\}_{a,x}\bigr) 
    &:= \max_{\{H_{a|x}\}_{a,x} \in \mathbf{H}^{(\rho)}_{+}} 
    \frac{Q_c\bigl(\{\rho_{a|x}\}, \{H_{a|x}\}\bigr)}
         {\max_{\{\sigma_{a|x}\} \in \mathcal{L}_{\rho}} 
         Q_c\bigl(\{\sigma_{a|x}\}, \{H_{a|x}\}\bigr)}\nonumber\\
    &> 1 + \mathcal{R}\bigl(\{\rho_{a|x}\}_{a,x}\bigr)\label{main_thm_equation},
\end{align}
where $\mathcal{R}(\{\rho_{a|x}\}_{a,x})$ denotes the steerability robustness, and $\mathcal{L}_{\rho}$ is defined in Eq.~\eqref{defn_Lrho}. The maximization is over all Hamiltonian tuples $\{H_{a|x}\}_{a,x} \in \mathbf{H}^{(\rho)}_{+}$ that result in withdrawal of heat from the bath, i.e.,
\begin{align}
    Q_c\bigl(\{\rho_{a|x}\}, \{H_{a|x}\}\bigr) \geq 0, \quad 
    \max_{\{\sigma_{a|x}\} \in \mathcal{L}_{\rho}}Q_c\bigl(\{\sigma_{a|x}\}, \{H_{a|x}\}\bigr) \geq 0.
\end{align}
\end{thm}
We prove Theorem~\ref{Robustness_assemblage} in Sec.~II of the Supplemental Material \cite{SM}, where we also derive a tighter temperature-dependent bound. In Sec.~III, we show that this bound decreases with increasing temperature, implying a prominent quantum advantage at lower bath temperatures compared to higher bath temperatures.
\blk

As established earlier, the steerability robustness 
\(\mathcal{R}\bigl(\{\rho_{a|x}\}_{a,x}\bigr)\) of any steerable assemblage 
\(\{\rho_{a|x}\}_{a,x}\) is strictly positive, which implies 
\(\xi_{\max}(\{\rho_{a|x}\}_{a,x}) > 1\). Consequently, Theorem~\ref{Robustness_assemblage} ensures the existence of a cooling task—i.e., a suitable choice of Hamiltonians—in which a steerable assemblage used as a resource outperforms any unsteerable assemblage in the set \(\mathcal{L}_{\rho}\). Remarkably, the maximal advantage in such a task is lower bounded by the steerability robustness. This provides a clear operational interpretation of steerability robustness, a geometric measure of steerability, in terms of a thermodynamic task.

Furthermore, the quantity \(\xi_{\max}(\{\rho_{a|x}\}_{a,x})\) itself serves as a witness of steerability and can thus be interpreted as a \emph{thermodynamic witness}. Indeed, if an assemblage \(\{\rho_{a|x}\}\) is unsteerable, then by definition \(\{\rho_{a|x}\} \in \mathcal{L}_{\rho}\) (see Eq.~\eqref{Eq_Fixed_weight_eqvt}). Since the denominator in Eq.~\eqref{main_thm_equation} involves a maximization over all assemblages in \(\mathcal{L}_{\rho}\), it follows that any unsteerable assemblage must satisfy \(\xi_{\max}(\{\rho_{a|x}\}) \leq 1\). Therefore, the condition \(\xi_{\max}(\{\rho_{a|x}\}) > 1\) provides a necessary and sufficient criterion for certifying steerability.

Next, we consider a class of steerable assemblages motivated by the connection between steerability and measurement incompatibility. As shown in Ref.~\cite{Uolaonetoone2015}, steering can be equivalently formulated as a joint measurability problem. We focus on assemblages generated from mutually unbiased bases (MUBs), which exhibit maximal incompatibility and hence maximal steering robustness. Using Eq.~\eqref{main_thm_equation}, this suggests a significant quantum advantage in cooling.

To establish this, we consider the following assemblage:
\begin{align}
    \rho_{a|x}^{\mathrm{isotropic}}(\eta)
    &:= \eta\, \frac{1}{d}\ketbra{\phi^{a}_{x}}
    + (1 - \eta)\,\frac{\mathbb{I}}{d^2},
    \quad 0 \leq \eta \leq 1,
    \label{MUB_assemblge}
\end{align}
where $a \in \{0,\ldots,d-1\}$ and $x \in \{0,\ldots,n\}$.  \cblue $\rho_{a|x}^{\mathrm{isotropic}}(\eta)$ can be interpreted as a convex mixture of assemblages arising from mutually unbiased bases (MUB) and the maximally mixed assemblage, with the mixing weight parametrized by $\eta$. 
An assemblage of the form $\{\rho_{a|x}^{\mathrm{isotropic}}(\eta)\}_{a,x}$ can be prepared if Alice and Bob share an isotropic entangled state \cite{HorodeckiIsotropic}, i.e., a state obtained as a convex mixture of a bipartite maximally entangled state and the maximally mixed state with the same weight $\eta$ (It represents a maximally entangled state affected by a depolarizing channel~\cite{whitenoise1,whitenoise2}): \blk
\begin{equation}
\rho^{\rm isotropic}_{AB}(\eta)
    = \eta\left(\frac{1}{d}\sum_{i,j=0}^{d-1}
    \ketbra{i}{j}_{A}\otimes\ketbra{i}{j}_{B}\right)
    + (1-\eta)\frac{\mathbb{I}_{AB}}{d^2},
\end{equation}
and Alice performs a measurement described by the projectors
\begin{equation}\label{MUB_measurement_assemblage}
    \left\{\ketbra{\phi^{*a}_{x}}{\phi^{*a}_{x}}\right\}_{a=0}^{d-1},
    \qquad 
    \ket{\phi^{*a}_{x}} = \sum_{i=0}^{d-1} \braket{\phi^a_x}{i}\ket{i}.
\end{equation}
Here, $\ket{\phi^a_x}$ denotes the $a$-th vector of the mutually unbiased basis (MUB) indexed by $x$, satisfying
\[
\bigl|\braket{\phi^a_x}{\phi^b_y}\bigr| =
\begin{cases}
\delta_{ab}, & x = y,\\[2mm]
1/\sqrt{d}, & x \neq y.
\end{cases}
\]
Note that for $\eta = 1$, the state $\rho^{\rm isotropic}_{AB}(\eta)$ becomes maximally entangled, and the assemblage $\rho_{a|x}^{\mathrm{isotropic}}(\eta)$ in Eq.~\eqref{MUB_assemblge} reduces to
\begin{equation}\label{MUB_assemblage_from_isotropic}
    \rho^{\mathrm{isotropic}}_{a|x}(\eta=1) =(d^{-1})\ketbra{\phi^a_x}=:\rho_{a|x}^{\mathrm{max.\; ent.}}.
\end{equation}

\begin{figure}[t]
    \centering
    \includegraphics[width=5.5 cm]{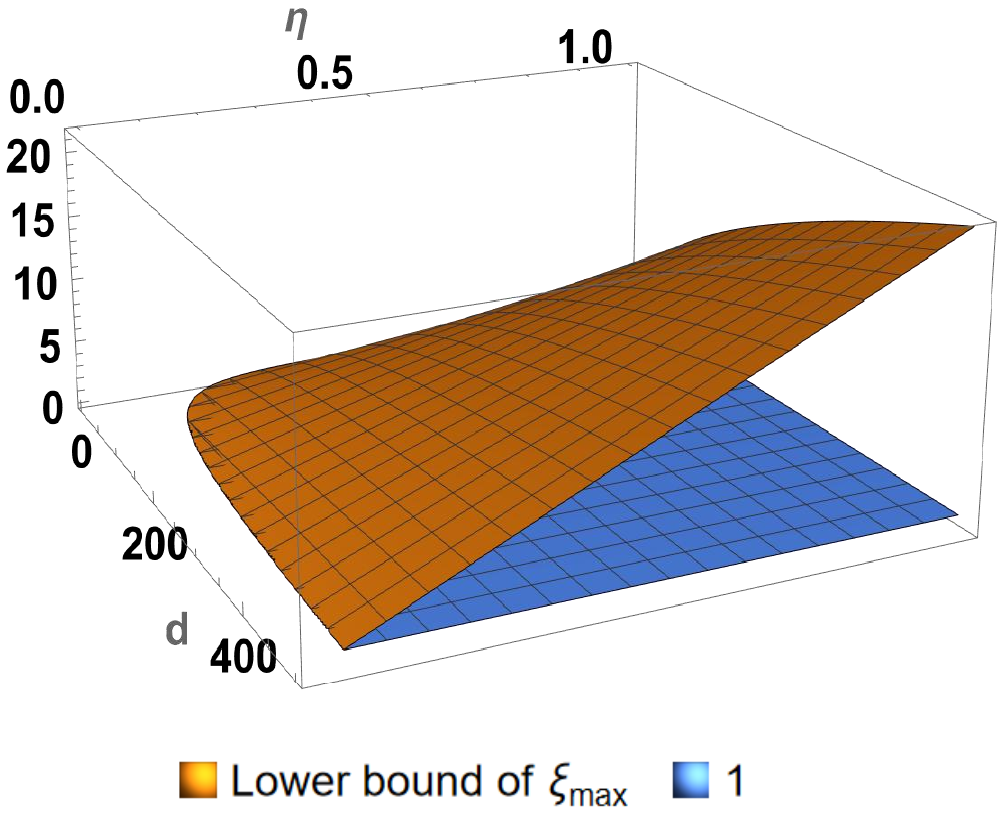}
\caption{\label{fig:quadratic_advantage} We plot the lower bound of 
$\xi_{\max}\!\left(\left\{\rho_{a|x}^{\mathrm{isotropic}}(\eta)\right\}_{a,x}\right)$, 
given in Eq.~\eqref{ximax_ineq}, as a function of $d$ and $\eta$ (orange surface) for 
$2 \leq d \leq 500$ and 
$\left(\sqrt{d}+\frac{1}{d+\sqrt{d}+1}\right)^{-1} < \eta \leq 1$. The surface lies strictly above the plane $z=1$ (blue), implying 
$\xi_{\max} > 1$ throughout this regime and thus confirming a quantum advantage, which increases with the dimension $d$.  
} 
\end{figure}

We emphasize that in any Hilbert space of dimension \(d = p^n\), where \(p\) is a prime and \(n\) is a positive integer, a complete set of \(d+1\) MUBs always exists. In such dimensions, we can always take the number of MUBs to be \(n = d+1\) \cite{WOOTTERS1989363, Bandyopadhyay2002}. As shown in Sec. V of Supplementary material \cite{SM}, in such dimensions taking $n = d + 1$ gives
\begin{equation}
    \mathcal{R}\!\left(\left\{\rho_{a|x}^{\mathrm{isotropic}}(\eta)\right\}_{a,x}\right)
    \geq 
    \sqrt{d}\!\left(\frac{\sqrt{d}-1}{\sqrt{d}+1}\right)\eta
    - \left(\frac{d\sqrt{d}-1}{d(\sqrt{d}+1)}\right)(1-\eta),
\end{equation}
which further gives using the inequality in Eq.~\eqref{main_thm_equation},
\begin{equation}\label{ximax_ineq}
    \xi_{\max}\!\left(\left\{\rho_{a|x}^{\mathrm{isotropic}}(\eta)\right\}_{a,x}\right)
    \geq
    \frac{(d+1)\bigl(1+\eta(d-1)\bigr)}{d\,(1+\sqrt{d})}.
\end{equation}

For $\eta = 1$, 
we obtain $\xi_{\max}\!\left(\left\{\rho_{a|x}^{\mathrm{max.\;ent.}}\right\}_{a,x}\right)
    \geq 
    (d+1)/(\sqrt{d}+1)
    = \mathcal{O}(\sqrt{d})$.
This demonstrates that the quantum advantage in the cooling task—when using steerable assemblages $\left\{\rho_{a|x}^{\mathrm{max.\;ent.}}\right\}_{a,x}$ as a resource—increases with the system dimension. This scaling of advantage growing with dimension is sometimes referred to as an unbounded quantum advantage \cite{Rutkowski2015,Saha2024,Rout2025}. \cblue In the end matter, for $d=2$, we consider an example based on the nuclear spin of nitrogen-vacancy (NV) center and obtain a tighter lower bound than that in Eq.~\eqref{ximax_ineq} for $\left\{\rho_{a|x}^{\mathrm{max.\,ent.}}\right\}_{a,x}$, using a Hamiltonian tuple constructed from Pauli operators. In particular, we find that $\xi_{\max}\!\left(\left\{\rho_{a|x}^{\mathrm{max.\,ent.}}\right\}_{a,x}\right) \geq \sqrt{3}$, which demonstrates a significant increase in the amount of heat withdrawn from the bath using quantum steerable assemblages. 
\blk

For the assemblage $\left\{\rho_{a|x}^{\mathrm{isotropic}}(\eta)\right\}_{a,x}$, whether the advantage grows with the dimension depends explicitly on how the parameter $\eta$ scales with $d$. In particular, achieving an advantage that increases with the dimension requires
\begin{equation}\label{eta+condition}
    \eta > \left(\sqrt{d} + \frac{1}{\,d + \sqrt{d} + 1\,}\right)^{-1},
\end{equation}
as illustrated in Fig. \ref{fig:quadratic_advantage} . The derivation of this condition is provided in Sec. VI of the supplementary material \cite{SM}.
 

\textit{Discussion.---} We show that quantum steerable assemblages constitute a resource for cooling, strictly outperforming all unsteerable ones. For any steerable assemblage, there exists a cooling protocol that achieves higher performance than any classical (unsteerable) counterpart. Quantifying the advantage via the ratio of extracted heat in the quantum and classical settings, we prove that it is lower bounded by the steering robustness, thereby establishing a direct connection between thermodynamic performance and quantum steerability.


%
These results open promising directions in quantum thermodynamics. As efficient cooling is crucial for scalable quantum technologies, the advantage provided by quantum correlations could play an important role in their practical realization. Moreover, since our protocol relies only on quenching and thermalization—standard operations in microscopic thermal machines—it is experimentally accessible with current technologies. \cblue Platforms such as superconducting qubits~\cite{Aamir2025}, trapped ions~\cite{Singer}, photoluminescent cooling in solid state devices \cite{Ghonge2024Photoluminescent} and NV centers~\cite{Steering_Engines_expt}, where tunable Hamiltonians and steerability have been demonstrated, provide natural testbeds for our proposal. \blk The framework also offers a route to explore connections between steering and non-Markovianity in quantum dynamics~\cite{ThomasGhosh}. Moreover, our techniques can be applied to achieve a quantum thermodynamic advantage using more general quantum resources, such as coherence and magic \cite{garciapintos2026robustnessthermodynamiccurrencywork}. 

\textit{Acknowledgments.---}We thank Shubhayan Sarkar, Chung-Yun Hsieh, Ryuji Takagi and Andreas Winter for their valuable comments. This material is based upon work supported by the U.S. Department of Energy, Office of Science, Accelerated Research in Quantum Computing, Fundamental Algorithmic Research toward Quantum Utility (FAR-Qu), and Fundamental Algorithmic Research in Quantum Computing (FAR-QC). We also acknowledge support from the Beyond Moore’s Law project of the Advanced Simulation and Computing Program at LANL managed by Triad National Security, LLC, for the National Nuclear Security Administration of the U.S. DOE under contract 89233218CNA000001.

\bibliography{main}

\newpage
\appendix
\cblue
\emph{Cooling advantages in Nitrogen-vacancy (NV) center:} To explore the cooling advantage in a realistic setting, we consider a system of qubits realized using nuclear spins in nitrogen–vacancy (NV) centers in diamond at room temperature $T=300 K$, for which
\begin{equation}\label{beta_NV_center}
    \beta=(k_BT)^{-1}=38.7 \;\text{eV}^{-1},
\end{equation}
where $k_B=8.62\times 10^{-5} \text{ eV/K}$ is the Boltzmann constant. 
The NV nuclear spin’s energy scale typically lies between 0.1MHz to 10 MHz. 

Consider a cooling task where Bob aims to withdraw heat from the bath using the following steerable assemblages $\rho_{a|x}$ as fuel, 
\begin{align}\label{Pauli_assemblages}
    &\rho_{0|0}^{\text{max. ent.}}=\frac{1}{2}\ketbra{0}{0}\quad\quad \rho_{1|0}^{\text{max. ent.}}=\frac{1}{2}\ketbra{1}{1},\\
    &\rho_{0|1}^{\text{max. ent.}}=\frac{1}{2}\ketbra{+}{+}\quad\quad \rho_{1|1}^{\text{max. ent.}}=\frac{1}{2}\ketbra{-}{-},\\
    &\rho_{0|2}^{\text{max. ent.}}=\frac{1}{2}\ketbra{+y}{+y}\quad\quad \rho_{1|2}^{\text{max. ent.}}=\frac{1}{2}\ketbra{-y}{-y},
\end{align}
where $a\in\{0,1\}$ and $x\in\{0,1,2\}$. 
Here, $\ket{0}$ and $\ket{1}$ denote the eigenstates of $Z$ with eigenvalues $+1$ and $-1$, respectively. Similarly, $\ket{+}$ and $\ket{-}$ are the eigenstates of $X$ with eigenvalues $+1$ and $-1$, while $\ket{+y}$ and $\ket{-y}$ represent the eigenstates of $Y$ corresponding to eigenvalues $+1$ and $-1$, respectively. $X$, $Y$ and $Z$ are the Pauli matrices. Such an assemblage can arise if Alice and Bob share a maximally entangled two-qubit state
\begin{equation}
\rho_{AB}^{\text{max. ent.}}=\frac{1}{2}\sum_{i,j=0}^{1}
\ketbra{i}{j}_{A}\otimes\ketbra{i}{j}_{B}.
\end{equation}
The assemblage is obtained when Alice performs measurements of $Z$, $X$, and $Y$ on her subsystem, respectively. 

Instead of determining the optimal Hamiltonian tuple in \(\textbf{H}_{+}^{(\rho)}\) that maximizes the quantum advantage \(\xi(\{\rho_{a|x}\}_{a,x})\) for the assemblage given in Eq.~\eqref{Pauli_assemblages}, we consider, for experimental feasibility, a Hamiltonian tuple \(\{H_{a|x}\}_{a,x}\) indexed by \(a \in \{0,1\}\) and \(x \in \{0,1,2\}\), constructed from the Pauli operators \(Z\), \(X\), and \(Y\). 
The Hamiltonians are as follows:
\begin{align}
    H_{a|0} &=   \epsilon \, (-1)^{(1-a)}Z,\label{Hamiltonian_qubit}\\
    H_{a|1} &=  \epsilon \,  (-1)^{(1-a)}X,\label{Hamiltonian_qubit2}\\
    H_{a|2} &= \epsilon \,  (-1)^{(1-a)}Y\label{Hamiltonian_qubit3},
\end{align}
where $\epsilon$ is a parameter that sets the energy scale of the Hamiltonian. This choice of Hamiltonian is motivated from the steering inequality presented in Eq. (65) of Ref. \cite{CavalcantiSteering} (See also Ref. \cite{PuseySteering}). Since we focus on NV nuclear spins at room temperature, we choose $\epsilon =  4.14 \times 10^{-9}\;\text{eV}$, which means
\begin{equation}\label{energy_scale}
\|H_{a|x}\|=\epsilon = 4.14 \times 10^{-9}\ \text{eV} \ (\approx 1\ \text{MHz}),
\end{equation}
which lies within the typical energy scale of NV nuclear spins.

Then, the amount of heat withdrawn from the bath using the Hamiltonians in Eqs.~\eqref{Hamiltonian_qubit}-\eqref{Hamiltonian_qubit3} is given by Eq. \eqref{Eq_Avg_heat_withdrawn} of the main text  
\cblue
\begin{align}\label{sub1}
    &Q_c\left(\{\rho_{a|x}^{\text{max. ent}}\},\{H_{a|x}\}\right)\nonumber\\=&\frac{1}{3}\sum_{x=0}^{2}\sum_{a=0}^{1}\left[\Tr\left(H_{a|x}\gamma_{a|x}\right)-\Tr\left(H_{a|x}\rho_{a|x}^{\text{max. ent.}}\right)\right],
\end{align}
where $\{\gamma_{a|x}\}_{a,x}$ is the collection of sub-normalized Gibbs state i.e.,
\begin{equation}\label{Gibbs_assemblage_supp}
    \{\gamma_{a|x}\}_{a,x}=\left\{\frac{1}{2}\times \frac{e^{-\beta H_{a|x}}}{\Tr\left(e^{-\beta H_{a|x}}\right)}\right\}_{a,x}.
\end{equation}
The pre-factor $1/2$ appearing in Eq.~\eqref{Gibbs_assemblage_supp} arises from the fact that $\Tr(\gamma_{a|x})=\Tr(\rho^{\text{max.\,ent.}}_{a|x})=1/2$, as can be seen from Eq. \eqref{Eq_assemblages} of the main text. Then, one can easily compute
\begin{align}
    \Tr \left(H_{a|x}\rho^{\text{max. ent.}}_{a|x}\right) = -\frac{\epsilon}{2}\;\;\text{and}\;\;\Tr \left(H_{a|x}\gamma_{a|x}\right) &= -\frac{\epsilon\tanh(\beta\epsilon)}{2}\label{gi1},
\end{align}
for any $a\in\{0,1\}$ and $x\in\{0,1,2\}$. $Q_c\left(\{\rho_{a|x}^{\text{max. ent}}\},\{H_{a|x}\}\right)$ is obtained by substituting Eq.~\eqref{gi1} into Eq.~\eqref{sub1}:
\begin{align}\label{f1ineq}
    Q_c\left(\{\rho_{a|x}^{\text{max. ent}}\},\{H_{a|x}\}\right)=\epsilon\left[1-\tanh(\beta\epsilon)\right].
\end{align}
As $\epsilon$ is order of $\sim 10^{-9}$ we have $\tanh(\epsilon)\ll 1$ and  
\begin{equation}\label{ste1}
    Q_c\left(\{\rho_{a|x}^{\text{max. ent}}\},\{H_{a|x}\}\right)\simeq \epsilon = 4.14 \times 10^{-9} \text{eV}\;(\approx 1 \text{MHz}),
\end{equation}
which lies within NV nuclear spin energy scale. 

Next, we demonstrate that the Hamiltonian tuple defined in 
Eqs.~\eqref{Hamiltonian_qubit}--\eqref{Hamiltonian_qubit3} belongs to the set 
$\mathbf{H}_{+}^{(\rho)}$. From Eq.~\eqref{ste1}, it follows that $Q_c\!\left(\{\rho_{a|x}^{\text{max. ent}}\},\{H_{a|x}\}\right) \geq 0$ for the inverse temperature $\beta$ in Eq.~\eqref{beta_NV_center} 
and the energy parameter $\epsilon$ in Eq.~\eqref{energy_scale}. 
Next, we show that
\begin{equation}\label{abov}
    \max_{\{\sigma_{a|x}\}\in \mathcal{L}_{\rho}} 
    Q_c\!\left(\{\sigma_{a|x}\},\{H_{a|x}\}\right) \geq 0,
\end{equation}
by lower bounding left hand side of the above inequality in Eq. \eqref{abov} via taking the following sub-optimal assemblage in $\mathcal{L}_{\rho}$: 
\begin{align}
    \sigma^{\text{sub-optimal}}_{0|0} &= \frac{1}{2}\ketbra{0}\quad\sigma^{\text{sub-optimal}}_{1|0} = \frac{1}{2}\ketbra{1}{1},\label{LHS_qubit_assemblage1}\\\sigma^{\text{sub-optimal}}_{0|1} &= \frac{1}{4}\mathbb{I}\quad\sigma^{\text{sub-optimal}}_{1|1} = \frac{1}{4}\mathbb{I},\label{LHS_qubit_assemblage2}\\\sigma^{\text{sub-optimal}}_{0|2} &= \frac{1}{2}\ketbra{1}\quad\sigma^{\text{sub-optimal}}_{1|2} = \frac{1}{2}\ketbra{0}\label{LHS_qubit_assemblage3}.
\end{align}
One can verify that the assemblage in Eqs.~\eqref{LHS_qubit_assemblage1}--\eqref{LHS_qubit_assemblage3} belongs to $\mathcal{L}_{\rho}$ by checking $\Tr(\sigma^{\text{sub-optimal}}_{a|x})=\Tr(\rho^{\text{max. ent.}}_{a|x})=\frac{1}{2}$ as well as its unsteerability [See Eq.~\eqref{defn_Lrho} in main text]. To verify the assemblage given in Eq. \eqref{LHS_qubit_assemblage1}--Eq.~\eqref{LHS_qubit_assemblage3} is unsteerable, we take $\lambda\in\{0,1\}$ such that
\begin{align}
  p(\lambda=0)&=p(\lambda=1)=\frac{1}{2} \quad\quad \text{and} \nonumber\\\sigma_{\lambda=0}&=\ketbra{0}\quad;\quad\sigma_{\lambda=1}=\ketbra{1},
\end{align}
and conditional probability $p(a|x,\lambda)$ with
\begin{align}
    p(0|0,\lambda=0)&=1\quad p(1|0,\lambda=0)=0 \\ p(0|1,\lambda=0)&=\frac{1}{2}\quad p(1|1,\lambda=0)=\frac{1}{2}\\p(0|2,\lambda=0)&=0\quad p(1|2,\lambda=0)=1\\ p(0|0,\lambda=1)&=0\quad p(1|0,\lambda=1)=1 \\ p(0|1,\lambda=1)&=\frac{1}{2}\quad p(1|1,\lambda=1)=\frac{1}{2}\\p(0|2,\lambda=1)&=1\quad p(1|2,\lambda=1)=0
\end{align}
in the definition of unsteerable assemblage given in Eq. \eqref{Eq_LHS_decomposition}. Then,
\begin{align}\label{ste9}
    &\max_{\{\sigma_{a|x}\}\in\mathcal{L}_{\rho}}Q_c\left(\{\sigma_{a|x}\},\{H_{a|x}\}\right)\geq Q_c\left(\left\{ \sigma^{\text{sub-optimal}}_{a|x}\right\}_{a,x},\{H_{a|x}\}\right)\nonumber\\&= \frac{1}{3}\sum_{x=0}^{2}\sum_{a=0}^{1}\left[\Tr\left(H_{a|x}\gamma_{a|x}\right)-\Tr\left(H_{a|x}\sigma^{\text{sub-optimal}}_{a|x}\right)\right]\nonumber\\
    &=\epsilon\left[\frac{1}{3}-\tanh(\beta\epsilon)\right]\geq 0,
\end{align}
where the inequality in Eq.~\eqref{ste9} follows by substituting $\epsilon$ in Eq.~\eqref{energy_scale} and the inverse temperature $\beta$ in Eq.~\eqref{beta_NV_center}. From Eqs.~\eqref{ste1} and~\eqref{ste9}, it follows that the Hamiltonian tuple 
$\{H_{a|x}\}_{a,x}$ defined in Eq.~\eqref{Hamiltonian_qubit}--\eqref{Hamiltonian_qubit3} 
belongs to $\mathbf{H}^{(\rho)}_{+}$. Moreover, Eq.~\eqref{ste9} holds not only for the 
specific inverse temperature in Eq.~\eqref{beta_NV_center}, but for the entire range:
\begin{equation}\label{beta_bound}
    0 \leq \beta \leq \frac{\ln 2}{2\epsilon} = 8.37 \times 10^7\ \text{eV}^{-1},
\end{equation}
for the energy scale given in Eq.~\eqref{energy_scale}. Hence, for all inverse temperatures within the 
range specified in Eq.~\eqref{beta_bound}, the Hamiltonian tuple 
$\{H_{a|x}\}_{a,x}$ in Eq.~\eqref{Hamiltonian_qubit}--\eqref{Hamiltonian_qubit3}  lies in $\mathbf{H}^{(\rho)}_{+}$. Substituting $\eta=1$ and $d=2$ in the Eq. \eqref{ximax_ineq} gives
\begin{equation}\label{ineqximaxendmatt}
     \xi_{\max}\!\left(\left\{\rho_{a|x}^{\mathrm{max.\;ent.}}\right\}_{a,x}\right)\geq \frac{3}{\sqrt{2}+1}.
\end{equation}

We can provide a lower bound tighter than Eq.~\eqref{ineqximaxendmatt}. To achieve this, we provide an upper bound on the heat that can be withdrawn from the bath using any unsteerable assemblages $\{\sigma_{a|x}\}_{a,x}\in\mathcal{L}_{\rho}$  as follows:
\cblue
\begin{align}\label{inequality_heat_supp_mat}
    &\max_{\{\sigma_{a|x}\}\in\mathcal{L}_{\rho}}Q_c\left(\{\sigma_{a|x}\},\{H_{a|x}\}\right)\leq  \max_{\{\sigma_{a|x}\}\in\mathcal{L}}Q_c\left(\{\sigma_{a|x}\},\{H_{a|x}\}\right)\nonumber\\=&\frac{1}{3}\max_{\{\sigma_{a|x}\}\in\mathcal{L}}\sum_{x=0}^{2}\sum_{a=0}^{1}\left[\Tr\left(H_{a|x}\gamma_{a|x}\right)-\Tr\left(H_{a|x}\sigma_{a|x}\right)\right]\nonumber\\
    =&\frac{1}{3}\sum_{x=0}^{2}\sum_{a=0}^{1}\Tr\left(H_{a|x}\gamma_{a|x}\right)-\min_{\{\sigma_{a|x}\}\in\mathcal{L}}\frac{1}{3}\sum_{x=0}^2\sum_{a=0}^1\Tr\left(H_{a|x}\sigma_{a|x}\right)\nonumber
     \\=&-\epsilon \tanh(\beta\epsilon)+ \epsilon\max_{\{\sigma_{a|x}\}\in\mathcal{L}}\frac{1}{3} \sum_{x=0}^2\sum_{a=0}^1\Tr\left(R_{a|x}\sigma_{a|x}\right)\nonumber\\     
     \leq&\epsilon\left[\frac{1}{\sqrt{3}}-\tanh(\beta\epsilon)\right],
\end{align}
where we define the operator $R_{a|x}:=-\epsilon^{-1}H_{a|x}$ for $a\in\{0,1\}$ and $x\in\{0,1,2\}$ i.e., $ R_{0|0}=-R_{1|0}=Z,\quad R_{0|1}=-R_{1|1}=X,\quad R_{0|2}=-R_{1|2}=Y.$

To write the first inequality in Eq. \eqref{inequality_heat_supp_mat} we simply use the fact that $\mathcal{L}_{\rho}$ is contained in the set of all unsteerable assemblages $\mathcal{L}$. Here, to write the final inequality we have used a result regarding unsteerable assemblages from Ref. \cite{PuseySteering} (originally derived from \cite{CavalcantiSteering}) 
\begin{equation}
    \sum_{x=0}^2\sum_{a=0}^{1} \Tr\left(\sigma_{a|x}R_{a|x}\right)\leq \sqrt{3}\quad \text{for any}\quad \{\sigma_{a|x}\}\in\mathcal{L},
\end{equation}

As $\epsilon$ is order of $\sim 10^{-9}$ thus $\tanh(\beta\epsilon)\ll 1/\sqrt{3}$, we have 
\begin{equation}\label{ste2}
   \max_{\sigma_{a|x}\in\mathcal{L}}Q_c\left(\{\rho_{a|x}^{\text{max. ent}}\},\{H_{a|x}\}\right)\lesssim \frac{\epsilon}{\sqrt{3}} = 2.39 \times 10^{-9} \text{eV}
\end{equation}
which approximately $0.578\; \text{MHz}$ hence still lies within NV nuclear spin energy scale. Using Eq. \eqref{ste1} and Eq. \eqref{ste2} we have
\begin{align}
  \xi_{\max}\!\left(\left\{\rho_{a|x}^{\mathrm{max.\;ent.}}\right\}_{a,x}\right)&\geq \frac{Q_c\left(\{\rho_{a|x}^{\text{max. ent}}\},\{H_{a|x}\}\right)}{\max_{\{\sigma_{a|x}\}\in\mathcal{L}_{\rho}}Q_c\left(\{\sigma_{a|x}\},\{H_{a|x}\}\right)}\label{ba}\\ 
  &\geq \frac{1-\tanh(\beta\epsilon)}{\frac{1}{\sqrt{3}}-\tanh(\beta\epsilon)}\label{f2ineq}
  =\frac{4.14\times 10^{-9}}{2.39\times 10^{-9}} \simeq \sqrt{3}.
\end{align}
This implies that steerable quantum assemblages are at least a factor of $\sqrt{3} \approx 1.732$ more efficient than unsteerable assemblages in withdrawing heat from the bath. Here, the inequality in Eq. \eqref{ba} follows from choosing a sub-optimal Hamiltonian tuple, and the second inequality in Eq. \eqref{f2ineq} is obtained from Eq.~\eqref{f1ineq} and \eqref{inequality_heat_supp_mat}, and the equality in Eq. \eqref{f2ineq} follows from Eq. \eqref{ste2}.

Next, we show that the lower bound \(\xi_{\max}\!\left(\left\{\rho_{a|x}^{\mathrm{max.\;ent.}}\right\}_{a,x}\right)\) obtained in Eq.~\eqref{f2ineq} is an increasing function of the inverse temperature \(\beta\) where $\beta$ satisfy bound given in Eq. \eqref{beta_bound} [This is required to ensure the Hamiltonian tuple considered in Eq. \eqref{Hamiltonian_qubit}--Eq. \eqref{Hamiltonian_qubit3} is contained in $\textbf{H}^{(\rho)}_{+}$]. To this end, we differentiate the bound in Eq.~\eqref{f2ineq} with respect to \(\beta\):
\begin{align}
    \frac{d}{d\beta}\left(\frac{1-\tanh(\beta\epsilon)}{\frac{1}{\sqrt{3}}-\tanh(\beta\epsilon)}\right)=\epsilon\left(\frac{\sqrt{3}-1}{\sqrt{3}}\right)\frac{3\sech^2(\beta\epsilon)}{\left(1-\sqrt{3}\tanh(\beta\epsilon)\right)^2}\geq 0,
\end{align}
which is non-negative. This shows that the lower bound in Eq.~\eqref{f2ineq} increases monotonically with $\beta$ (i.e., decreases with temperature), implying that the heat-withdrawal advantage is enhanced at lower temperatures. We revisit and generalize this in Sec.~III of the Supplemental Material \cite{SM}.
\blk
\newpage
\onecolumngrid
\section*{Supplementary material: All steerable quantum correlations can provide thermodynamic advantages in cooling}

\cblue
\section{A useful result regarding the unsteerable assemblages}
\begin{prop}\label{prop_steerable_assemblage}
For a given assemblage $\{\rho_{a|x}\}_{a,x}$ with $\Tr(\rho_{a|x}) = p(a|x)$ for all $a,x$, every assemblage of the form $\{p(a|x)\,\hat{\varrho}\}_{a,x}$, where $\hat{\varrho}$ is an arbitrary normalized quantum state, belongs to $\mathcal{L}_{\rho}$.
\end{prop}
\begin{proof}
Recall that an assemblages $\{\sigma_{a|x}\}_{a,x}$ said to be unsteerable iff it can be written in the following manner:
\begin{equation}\label{Eq_LHS_decomposition_supp}
    \sigma_{a|x}=\int d\lambda\, p(\lambda)\, p(a|x,\lambda)\, \sigma_\lambda,
    \qquad \forall\, a,x.
\end{equation}
We set \( p(\lambda)=1 \), \( p(a|x,\lambda)=p(a|x) \) for all \( a,x \), and \( \sigma_{\lambda}=\hat{\varrho} \). This verifies that the assemblage of the form $\{p(a|x)\hat{\varrho}\}_{a,x}$, is unsteerable. Moreover, since $\hat{\varrho}$ is a normalized state we have $\forall a,x;\ \Tr(p(a|x)\hat{\varrho}) = p(a|x) = \Tr(\rho_{a|x})$. This implies that $\{p(a|x)\hat{\varrho}\}_{a,x} \in \mathcal{L}_{\rho}$.     
\end{proof}
\section{Proof of theorem 1}
For a given steerable assemblage $\{\rho_{a|x}\}_{a,x}$ with $\Tr(\rho_{a|x}) = p(a|x)$, one can calculate the steerability robustness (see Refs.~\cite{Piani_subchannel,SteeringRPP}) using the following semidefinite program (SDP):

\begin{equation}\label{Dual_robustness}
\begin{aligned}
1+\mathcal{R}(\{\rho_{a|x}\}_{a,x})=\text{maximize} \quad & \sum_{a,x} \operatorname{Tr}\left[ F_{a|x} \rho_{a|x} \right] \\
\text{subject to} \quad & \sum_{a,x} D(a|x,\lambda) F_{a|x} \leq \mathbb{I}, \quad \forall \lambda, \\
& F_{a|x} \geq 0, \quad \forall a,x,
\end{aligned}
\end{equation}
where $D(a|x,\lambda)$ is the deterministic response function i.e., $D(a|x,\lambda)=\delta_{a,\lambda(x)}$. We denote $\{F^{*}_{a|x}\}_{a,x}$ as the solution of the SDP given in Eq. \eqref{Dual_robustness} which means
\begin{equation}\label{equality_opt}
   \sum_{a,x} \operatorname{Tr}\left[ F^{*}_{a|x} \rho_{a|x} \right]= 1+\mathcal{R}(\{\rho_{a|x}\}_{a,x}).
\end{equation}
To prove Theorem 1, we divide the proof into two cases:
\begin{enumerate}
    \item Case 1: $\sum_{a,x} p(a|x)\, F^{*}_{a|x}  \not\propto \mathbb{I}$ i.e $\sum_{a,x} p(a|x)\, F^{*}_{a|x}$ is not fully degenerate.
    \item Case 2: $\sum_{a,x} p(a|x)\, F^{*}_{a|x} \propto \mathbb{I}$ i.e $\sum_{a,x} p(a|x)\, F^{*}_{a|x}$ is fully degenerate.
\end{enumerate}

\subsection{Proof for Case 1}
We first consider Case 1,  where $\sum_{a,x} p(a|x)\, F^{*}_{a|x}$ is not fully degenerate, i.e.,
\begin{equation}\label{Case1}
    \sum_{a,x} p(a|x)\, F^{*}_{a|x} \not\propto \mathbb{I}.
\end{equation}
Consider a Hamiltonian tuple of the form  
\begin{equation}\label{Hamiltonian_from_witness}
    \{H^{(\epsilon)}_{a|x}\} = \{-\epsilon\, F^{*}_{a|x}\}_{a,x} \quad\text{with $\epsilon>0$},
\end{equation}
which Bob quenches to after receiving the pair \((a,x)\) from Alice. Here the parameter $\epsilon$ encapsulates the strength of the quench.

We define the optimal unsteerable assemblage $\{\sigma^{*}_{a|x}\}_{a,x}\in \mathcal{L}_{\rho}$ associated with $\{F^{*}_{a|x}\}_{a,x}$ via the following relation:
\begin{align}\label{Lrhomax}
\max_{\{\sigma_{a|x}\} \in \mathcal{L}_{\rho}}
    \sum_{a,x}\Tr\bigl(F^{*}_{a|x}\sigma_{a|x}\bigr) 
    :=  \sum_{a,x}\Tr\bigl(F^{*}_{a|x}\sigma^{*}_{a|x}\bigr). 
\end{align}
Since $\{\sigma^{*}_{a|x}\}_{a,x}$ is an unsteerable assemblage, it holds that~\cite{SteeringRPP,SteeringRevModPhys}:
\begin{equation}\label{Eq_unsteerable_witness}
    \sum_{a,x}\Tr\bigl(F^{*}_{a|x}\sigma^{*}_{a|x}\bigr) \leq 1, 
    \quad\quad \forall\, a,x.
\end{equation}
We also define a collection of sub-normalized Gibbs states with respect to the Hamiltonian in Eq.~\eqref{Hamiltonian_from_witness} at bath's inverse temperature $\beta$:  
\begin{align}\label{Eq_Gibbs_Assemblages_pre}
    \gamma^{(\epsilon)}_{a|x}&:=p(a|x)\hat\gamma^{(\epsilon)}_{a|x}\quad\text{where}\nonumber\\ \;\; \hat\gamma^{(\epsilon)}_{a|x}&:=\frac{e^{-\beta H^{(\epsilon)}_{a|x}}}{\Tr\left(e^{-\beta H^{(\epsilon)}_{a|x}}\right)}=\frac{e^{\epsilon\beta  F^{*}_{a|x}}}{\Tr\left(e^{\epsilon \beta F^{*}_{a|x}}\right)}.
\end{align}

First, we show that one can choose a sufficiently small but positive  $\epsilon$ i.e., $\epsilon=\epsilon_{*}>0$  such that resulting Hamiltonian tuple $\{H_{a|x}^{*}\}_{a,x} := \{H_{a|x}^{(\epsilon_{*})}\}_{a,x}=\{-\epsilon_{*}F^{*}_{a|x}\}_{a,x} \in \textbf{H}^{(\rho)}_{+}$, 
which means
\begin{align}\label{stareqn}
    \max_{\{\sigma_{a|x}\} \in \mathcal{L}_{\rho}}Q_c\bigl(\{\sigma_{a|x}\}, \{H^{*}_{a|x}\}\bigr) \geq 0\quad \text{and}\quad Q_c\bigl(\{\rho_{a|x}\}, \{H^{*}_{a|x}\}\bigr) \geq 0. 
\end{align}
 
Let us proceed by showing that $\max_{\{\sigma_{a|x}\}\in\mathcal{L}_{\rho}}Q_c(\{\sigma_{a|x}\},\{H^{*}_{a|x}\})\geq 0$. For any $\epsilon>0$, we have
\begin{align}
    \max_{\{\sigma_{a|x}\}\in\mathcal{L}_{\rho}}Q_c(\{\sigma_{a|x}\},\{H^{(\epsilon)}_{a|x}\}) =& \max_{\{\sigma_{a|x}\}\in\mathcal{L}_{\rho}}\frac{1}{n}\sum_{a,x} \left[  \Tr \bigl( H^{(\epsilon)}_{a|x} \gamma^{(\epsilon)}_{a|x}\bigr) -\Tr \bigl( H^{(\epsilon)}_{a|x}\sigma_{a|x} \bigr)
         \right],\nonumber\\
    =&\frac{\epsilon}{n}\sum_{a,x} \Tr(-F^{*}_{a|x}\gamma^{(\epsilon)}_{a|x})-\min_{\{\sigma_{a|x}\}\in\mathcal{L}_{\rho}}\frac{\epsilon}{n}\sum_{a,x} \Tr(- F^{*}_{a|x}\sigma_{a|x}),\\     
    =& \max_{\{\sigma_{a|x}\}\in\mathcal{L}_{\rho}}\frac{\epsilon}{n}\sum_{a,x} \Tr( F^{*}_{a|x}\sigma_{a|x})-\frac{\epsilon}{n}\sum_{a,x} \Tr(F^{*}_{a|x}\gamma^{(\epsilon)}_{a|x}),\label{Eq_29}\\ 
    =& \frac{\epsilon}{n}\sum_{a,x}\left[\Tr\left(F^{*}_{a|x}\sigma^{*}_{a|x}\right)-\Tr( F^{*}_{a|x}\gamma^{(\epsilon)}_{a|x})\right]. \label{Eq_30}
\end{align}
To obtain the third equality in Eq.~\eqref{Eq_29}, we use that for any operator $A$,
\begin{equation}\label{identity}
    -\min_{X \in \mathcal{S}} \mathrm{Tr}(-A X)
    =
    \max_{X \in \mathcal{S}} \mathrm{Tr}(A X).
\end{equation}
The final equality in Eq.~\eqref{Eq_30} follows from Eq.~\eqref{Lrhomax}. Note that for $\epsilon = 0$, 
\begin{equation}\label{Eq_maximally_mixed_assemblage}
    \gamma^{(0)}_{a|x} = p(a|x)\frac{\mathbb{I}}{d},\quad\quad \forall\, a,x,
\end{equation}
which can be obtained directly from Eq. \eqref{Eq_Gibbs_Assemblages_pre}. Observe that the assemblage $\left\{p(a|x)\frac{\mathbb{I}}{d}\right\}_{a,x}$ belongs to the set $\mathcal{L}_{\rho}$, as follows from Proposition~\ref{prop_steerable_assemblage}. Thus, using Eq. \eqref{Eq_maximally_mixed_assemblage}, we obtain the following: 
\begin{align}\sum_{a,x}\Bigl[\Tr\!\left(F^{*}_{a|x}\sigma^{*}_{a|x}\right)
    - \Tr\!\left(F^{*}_{a|x}\gamma^{(0)}_{a|x}\right)\Bigr] =& \sum_{a,x}\left[\Tr \Big( F^{*}_{a|x}\sigma^{*}_{a|x}\Big)- p(a|x)\Tr \Big( F^{*}_{a|x}\frac{\mathbb{I}}{d}\Big)\right],\label{inequality_denominator23}\\
    \geq & \sum_{a,x}\left[p(a|x)\Tr \Big( F^{*}_{a|x}\ketbra{\psi}{\psi}\Big)- p(a|x)\Tr \Big( F^{*}_{a|x}\frac{\mathbb{I}}{d}\Big)\right],\label{inequality_denominator234}\\
    >&\;0, \label{inequality_denominator2345}
\end{align}
where $\ket{\psi}$ is the eigenvector corresponding to the largest eigenvalue of the operator $\sum_{a,x}p(a|x)F^{*}_{a|x}$. To write the equality in Eq. \eqref{inequality_denominator23} we simply use Eq. \eqref{Eq_maximally_mixed_assemblage}. The first inequality in Eq.~\eqref{inequality_denominator234} follows from choosing a suboptimal assemblage $\{p(a|x)\ketbra{\psi}\}_{a,x} \in \mathcal{L}_{\rho}$ (See Eq. \eqref{Lrhomax}), where the inclusion $\{p(a|x)\ketbra{\psi}\}_{a,x} \in \mathcal{L}_{\rho}$ follows from Proposition~\ref{prop_steerable_assemblage}. The strict inequality follows from the non-degeneracy of the operator \(\sum_{a,x} p(a|x)\, F^{*}_{a|x}\), as established in Eq.~\eqref{Case1}.

Since the inequality in Eq.~\eqref{inequality_denominator2345} holds strictly at $\epsilon=0$, and the quantity
\begin{equation}
\sum_{a,x}\Bigl[\Tr\!\left(F^{*}_{a|x}\sigma^{*}_{a|x}\right)
- \Tr\!\left(F^{*}_{a|x}\gamma^{(\epsilon)}_{a|x}\right)\Bigr],
\end{equation}
is continuous in $\epsilon$, it follows by continuity that there exists $\epsilon=\epsilon_{*} > 0$, sufficiently close to $\epsilon=0$, such that
\begin{align}\label{qua2}
    \sum_{a,x}\left[\Tr\!\left(F^{*}_{a|x}\sigma^{*}_{a|x}\right)
    - \Tr\!\left(F^{*}_{a|x}\gamma^{(\epsilon_{*})}_{a|x}\right)\right] > 0,
\end{align}
and employing Eq. \eqref{Eq_unsteerable_witness}, we also have an upper bound:
\begin{align}\label{qua234}
    1-\sum_{a,x}\Tr\!\left(F^{*}_{a|x}\gamma^{(\epsilon_{*})}_{a|x}\right)>& \sum_{a,x}\left[\Tr\!\left(F^{*}_{a|x}\sigma^{*}_{a|x}\right)
    - \Tr\!\left(F^{*}_{a|x}\gamma^{(\epsilon_{*})}_{a|x}\right)\right] > \;0.
\end{align}
 Thus, Eq.~\eqref{qua2} implies that the amount of heat withdrawn is as given in Eq.~\eqref{Eq_30}, using the Hamiltonian tuple $\{H^{*}_{a|x}\}_{a,x}=\{-\epsilon_{*}F^{*}_{a|x}\}_{a,x}$ and the $\{\sigma^{*}_{a|x}\}_{a,x}$ 
\begin{align}\label{qua3}
    &\max_{\{\sigma_{a|x}\}\in\mathcal{L}_{\rho}} Q_c(\{\sigma_{a|x}\},\{H^{*}_{a|x}\}) 
    = \frac{\epsilon_{*}}{n}\sum_{a,x}\Bigl[\Tr\!\left(F^{*}_{a|x}\sigma^{*}_{a|x}\right)
    - \Tr\!\left(F^{*}_{a|x}\gamma^{(\epsilon_{*})}_{a|x}\right)\Bigr] > 0.
\end{align}
Employing Eq. \eqref{Eq_unsteerable_witness}, we can obtain also the upper bound for $\max_{\{\sigma_{a|x}\}\in\mathcal{L}_{\rho}} Q_c(\{\sigma_{a|x}\},\{H^{*}_{a|x}\})$ starting from Eq. \eqref{qua3}
\begin{align}\label{final_ineq_31}
    \frac{\epsilon_{*}}{n}\left[1-\sum_{a,x} \Tr\!\left(F^{*}_{a|x}\gamma^{(\epsilon_{*})}_{a|x}\right)\right]&\geq \max_{\{\sigma_{a|x}\}\in\mathcal{L}_{\rho}} Q_c(\{\sigma_{a|x}\},\{H^{*}_{a|x}\}).
\end{align}

Next, we show that $Q_c(\{\rho_{a|x}\},\{H^{*}_{a|x}\})\geq 0$ as follows:
\begin{align}
    Q_c(\{\rho_{a|x}\},\{H^{*}_{a|x}\})&= \frac{1}{n}\sum_{a,x} \left[  \Tr \bigl( H^{*}_{a|x} \gamma^{(\epsilon_*)}_{a|x} \bigr) -\Tr \bigl( H^{*}_{a|x}\rho_{a|x} \bigr)\right],\nonumber\\
    &= \frac{\epsilon_{*}}{n}\sum_{a,x} \left[  \Tr \bigl( F^{*}_{a|x}\rho_{a|x} \bigr)-\Tr \bigl( F^{*}_{a|x} \gamma^{(\epsilon_*)}_{a|x} \bigr)\right],\nonumber\\
    &= \frac{\epsilon_{*}}{n} \left[  1+\mathcal{R}(\{\rho_{a|x}\}_{a,x})-\Tr \bigl( F^{*}_{a|x} \gamma^{(\epsilon_*)}_{a|x} \bigr)\right],\label{Qcstererable}\\
    &> 0\label{Qcstererable2}.
\end{align}
To write the final equality we use Eq. \eqref{equality_opt},
and to write the inequality, we use Eq.~\eqref{qua234} and the fact that $R(\{\rho_{a|x}\}_{a,x})>0$ for the steerable assemblage $\{\rho_{a|x}\}_{a,x}$. Therefore, Eqs.~\eqref{qua3} and~\eqref{Qcstererable2} imply that 
\begin{equation}\label{Hamiltonian_from_witness23}
    \{H^{*}_{a|x}\}_{a,x}=\{-\epsilon_{*}F^{*}_{a|x}\}_{a,x}\in \textbf{H}^{(\rho)}_{+}.
\end{equation}

Next, we prove that $ \xi_{\rm max}\left(\{\rho_{a|x}\}_{a,x}\right)\geq 1+\mathcal{R}(\{\rho_{a|x}\}_{a,x})$. We proceed by writing
\begin{align}
    \xi_{\rm max}\left(\{\rho_{a|x}\}_{a,x}\right) &\coloneqq \max_{\{H_{a|x}\}_{a,x}\in\textbf{H}^{(\rho)}_{+}} 
    \frac{Q_c\bigl(\{\rho_{a|x}\}, \{H_{a|x}\}\bigr)}
         {\max_{\{\sigma_{a|x}\} \in \mathcal{L}_{\rho}} 
         Q_c\bigl(\{\sigma_{a|x}\}, \{H_{a|x}\}\bigr)},  \nonumber \\
     &\geq 
    \frac{Q_c\bigl(\{\rho_{a|x}\}, \{H^{*}_{a|x}\}\bigr)}
         {\max_{\{\sigma_{a|x}\} \in \mathcal{L}_{\rho}} 
         Q_c\bigl(\{\sigma_{a|x}\}, \{H^{*}_{a|x}\}\bigr)},\nonumber\\    
    &\geq \frac{ \left[ 1+\mathcal{R}(\{\rho_{a|x}\}_{a,x}) - \sum_{a,x}\Tr \bigl( F^{*}_{a|x} \gamma^{(\epsilon_{*})}_{a|x} \bigr) \right]}{ \left[ 1 - \sum_{a,x}\Tr \bigl( F^{*}_{a|x} \gamma^{(\epsilon_{*})}_{a|x}\bigr) \right]},\label{last_step_app}\\
    &> 1+\mathcal{R}(\{\rho_{a|x}\}_{a,x})\label{inequality_last_lane},
\end{align}
where the first inequality follows from choosing a specific tuple of Hamiltonian as given in Eq.~\eqref{Hamiltonian_from_witness23}, rather than performing a maximization over all possible Hamiltonian tuples, and the second inequality is obtained by invoking Eq.~\eqref{Qcstererable} together with Eq.~\eqref{final_ineq_31}.  Note that the lower bound of $\xi_{\rm max}\left(\{\rho_{a|x}\}_{a,x}\right)$, as given in Eq.~\eqref{last_step_app}, explicitly depends on the temperature through the term $\sum_{a,x}\Tr \bigl( F^{*}_{a|x},\gamma^{(\epsilon)}_{a|x}\bigr)$. To derive the final inequality in Eq.~\eqref{inequality_last_lane}, we use the fact that
\begin{equation}
    \frac{x-z}{y-z}> \frac{x}{y}
    \quad \Longleftrightarrow \quad 
    y-z \geq 0 
    \quad \text{and} \quad 
    x > y,
\end{equation}
with the identifications $x = 1 + \mathcal{R}\bigl(\{\rho_{a|x}\}_{a,x}\bigr)$, $y = 1$ and
$z = \sum_{a,x}\Tr \bigl( F^{*}_{a|x} \gamma^{(\epsilon_{*})}_{a|x}\bigr)$. Since $\mathcal{R}\bigl(\{\rho_{a|x}\}_{a,x}\bigr) > 0$ for any steerable assemblage $\{\rho_{a|x}\}_{a,x}$, it follows that $x > y$, while $y - z \geq 0$ follows directly from Eq.~\eqref{qua234}. This completes the proof for case 1.

\subsection{Proof for case 2}
Next, we prove the claim for Case 2, namely when
\begin{equation}\label{case2}
    \sum_{a,x} p(a|x)\, F^{*}_{a|x} \propto \mathbb{I}
    \;\Longleftrightarrow\;
    \sum_{a,x} p(a|x)\, F^{*}_{a|x} = c\,\mathbb{I},
\end{equation}
for some constant \( c >0 \). For any assemblage of the form \( \{p(a|x)\hat{\varrho}\}_{a,x} \in \mathcal{L}_{\rho} \), using Eq.~\eqref{case2}, we obtain
\begin{equation}\label{cineq1}
    1 \geq \sum_{a,x} p(a|x)\, \Tr\!\left(F^{*}_{a|x}\varrho\right) = c,
\end{equation}
where the inequality follows from the fact that \( \{p(a|x)\varrho\} \) is an arbitrary assemblage in \( \mathcal{L}_{\rho} \) [see Eq.~\eqref{Eq_unsteerable_witness}].

Whenever Eq.~\eqref{case2} holds, the strict inequality in Eq.~\eqref{inequality_denominator2345} that arises due to the Hamiltonian tuple $\{H^{(\epsilon_*)}_{a|x}\}_{a,x}=\{-\epsilon_* F^{*}_{a|x}\}_{a,x}$ from Eq. \eqref{Hamiltonian_from_witness}, reduces to an equality for any assemblage of the form \( \{p(a|x)\hat{\varrho}\}_{a,x} \). Since the inequality is no longer strict, the continuity argument employed earlier does not apply in this case. Consequently, the previous proof breaks down if we choose the Hamiltonian of the form given in Eq.~\eqref{Hamiltonian_from_witness} under the condition that Eq.~\eqref{case2} holds. 

To study this case, we consider a perturbation of the previously-considered Hamiltonian tuple, with a perturbation parametrized by $\alpha$:
\begin{equation}\label{tildeHamiltonian}
    \{\tilde{H}^{(\epsilon)}_{a|x}(\alpha)\}_{a,x}=\{-\epsilon \tilde{F}^{*}_{a|x}(\alpha)\}_{a,x},
\end{equation}
where $\epsilon > 0$ and
\begin{equation}\label{Ftilde}
    \forall a,x, \quad \tilde{F}^{*}_{a|x}(\alpha):=\left[F^{*}_{a|x}+\alpha\left(\ketbra{\phi}-\frac{\mathbb{I}}{d}\right)\right].
\end{equation}
$\alpha>0$ and $\ket{\phi}$ is a fixed normalized quantum state. Here, $\alpha$ quantifies the strength of the perturbation generated by the operator $\left(\ketbra{\phi} - \frac{\mathbb{I}}{d}\right)$. We define a Gibbs state with respect to the Hamiltonian tuple defined in Eq.~\eqref{tildeHamiltonian} at the inverse temperature of the bath $\beta$:
\begin{equation}\label{collection_Gibbse}
    \tilde{\gamma}_{a|x}^{(\epsilon)}(\alpha):=p(a|x) \hat{\tilde{\gamma}}_{a|x}^{(\epsilon)}(\alpha)\quad\text{where}\quad \hat{\tilde{\gamma}}_{a|x}^{(\epsilon)}(\alpha):=\frac{e^{-\beta \tilde{H}^{(\epsilon)}_{a|x}(\alpha)}}{\Tr\left(e^{-\beta \tilde{H}^{(\epsilon)}_{a|x}(\alpha)}\right)}.
\end{equation}
Let us also define 
\begin{align}\label{tildesigmastar}
    \max_{\{\sigma_{a|x}\} \in \mathcal{L}_{\rho}}\Tr\left( \tilde{F}^{*}_{a|x}(\alpha)\sigma_{a|x}\right):=\Tr\left(\tilde{F}^{*}_{a|x}(\alpha)\tilde{\sigma}_{a|x}^{*}(\alpha)\right),
\end{align}
where we see that optimal assemblage $\tilde{\sigma}_{a|x}^{*}(\alpha)$ is a function of $\alpha$ and it satisfies
\begin{equation}\label{above_eqn_blah}
    \tilde{\sigma}_{a|x}^{*}(0) = \sigma_{a|x}^{*},
\end{equation}
because we see from Eq. \eqref{Ftilde} that $\tilde{F}^{*}_{a|x}(0)=F^{*}_{a|x}$ and employing Eq. \eqref{Lrhomax} we have Eq. \eqref{above_eqn_blah}. 

Firstly, we show that there exists a ball of radius $\delta_{1}$ around $(\epsilon,\alpha)=(0,0)$ where the the inequality 
\begin{equation}\label{desired_ineq}
    \xi_{\rm max}\left(\{\rho_{a|x}\}_{a,x}\right)> 1+\mathcal{R}(\{\rho_{a|x}\}_{a,x}),
\end{equation}
holds.
For any $\epsilon>0$ and $\alpha>0$ we have
\begin{align}\label{ximaxboundintermediate}
    \xi_{\rm max}\left(\{\rho_{a|x}\}_{a,x}\right) &\coloneqq \max_{\{H_{a|x}\}_{a,x}\in\textbf{H}^{(\rho)}_{+}} 
    \frac{Q_c\bigl(\{\rho_{a|x}\}, \{H_{a|x}\}\bigr)}
         {\max_{\{\sigma_{a|x}\} \in \mathcal{L}_{\rho}} 
         Q_c\bigl(\{\sigma_{a|x}\}, \{H_{a|x}\}\bigr)}\nonumber\\&\geq 
    \frac{Q_c\bigl(\{\rho_{a|x}\}, \{\tilde{H}^{(\epsilon)}_{a|x}(\alpha)\}\bigr)}
         {\max_{\{\sigma_{a|x}\} \in \mathcal{L}_{\rho}} 
         Q_c\bigl(\{\sigma_{a|x}\}, \{\tilde{H}^{(\epsilon)}_{a|x}(\alpha)\}\bigr)},
\end{align}
where the inequality follows from choosing a specific Hamiltonian tuple as given in Eq.~\eqref{tildeHamiltonian}, rather than performing a maximization over all possible Hamiltonian tuple in $\textbf{H}^{(\rho)}_{+}$. We will prove the desired inequality by simplifying the numerator and providing upper bound to the denominator in Eq. \eqref{ximaxboundintermediate}. 

Let us proceed by simplifying the numerator in Eq. \eqref{ximaxboundintermediate}. Using the expression for Hamiltonian tuple $\{H^{(\epsilon)}_{a|x}(\alpha)\}_{a,x}$ from Eq. \eqref{tildeHamiltonian} and collection of sub-normalized Gibbs state $\{\tilde{\gamma}_{a|x}^{(\epsilon)}(\alpha)\}_{a,x}$ from Eq. \eqref{collection_Gibbse}, we have for all $\epsilon>0$ and $\alpha>0$:
\begin{align}
    &Q_c(\{\rho_{a|x}\},\{H^{(\epsilon)}_{a|x}(\alpha)\}) = \frac{1}{n}\sum_{a,x} \left[  \Tr \bigl(H^{(\epsilon)}_{a|x}(\alpha) \tilde{\gamma}^{(\epsilon)}_{a|x}(\alpha) \bigr) -\Tr \bigl( H^{(\epsilon)}_{a|x}(\alpha)\rho_{a|x} \bigr)\right],\nonumber\\
   = & \frac{\epsilon}{n}\sum_{a,x} \Bigg[  \Tr \bigl( F^{*}_{a|x}\rho_{a|x} \bigr)+\alpha \Tr\left(\left(\ketbra{\phi}-\frac{\mathbb{I}}{d}\right)\rho_{a|x}\right)\Bigg]-\frac{\epsilon}{n} \sum_{a,x}\Bigg[  \Tr \bigl( F^{*}_{a|x}\rho_{a|x} \bigr)+\alpha \Tr\left(\left(\ketbra{\phi}-\frac{\mathbb{I}}{d}\right)\tilde{\gamma}_{a|x}^{(\epsilon)}(\alpha)\right)\Bigg],\\
    = & \frac{\epsilon}{n} \underbrace{\Bigg[  1+\mathcal{R}(\{\rho_{a|x}\}_{a,x})+\alpha \sum_{a,x}\Tr\left(\left(\ketbra{\phi}-\frac{\mathbb{I}}{d}\right)\rho_{a|x}\right)-\sum_{a,x}\Tr \left(\left[ F^{*}_{a|x}+\alpha \left(\ketbra{\phi}-\frac{\mathbb{I}}{d}\right)\right] \tilde{\gamma}_{a|x}^{(\epsilon)}(\alpha) \right)\Bigg]}_{:=f(\epsilon,\alpha)}\label{ineq_in_l41}=\frac{\epsilon}{n} f(\epsilon,\alpha),
\end{align}
where to write the first equality in Eq.~\eqref{ineq_in_l41} we use Eq.~\eqref{equality_opt}. Next, in a similar manner, by employing the expressions of $\{H^{(\epsilon)}_{a|x}(\alpha)\}_{a,x}$ from Eq.~\eqref{tildeHamiltonian} and $\{\tilde{\gamma}_{a|x}^{(\epsilon)}(\alpha)\}_{a,x}$ from Eq.~\eqref{collection_Gibbse}, we derive an upper bound on the denominator in Eq.~\eqref{ximaxboundintermediate} valid for all $\epsilon > 0$ and $\alpha > 0$, as follows:
\begin{align}
    &\max_{\{\sigma_{a|x}\}\in\mathcal{L}_{\rho}}Q_c(\{\sigma_{a|x}\},\{\tilde{H}^{(\epsilon)}_{a|x}(\alpha)\})= \max_{\{\sigma_{a|x}\}\in\mathcal{L}_{\rho}}\frac{1}{n}\sum_{a,x} \left[  \Tr \bigl( \tilde{H}^{(\epsilon)}_{a|x} (\alpha)\tilde{\gamma}^{(\epsilon)}_{a|x}(\alpha) \bigr) -\Tr \bigl( \tilde{H}^{(\epsilon)}_{a|x}(\alpha)\sigma_{a|x} \bigr)
         \right]\nonumber\\
    =&\frac{\epsilon}{n}\sum_{a,x} \Tr(-\tilde{F}^{*}_{a|x}(\alpha)\tilde{\gamma}^{(\epsilon)}_{a|x}(\alpha))-\min_{\{\sigma_{a|x}\}\in\mathcal{L}_{\rho}}\frac{\epsilon}{n}\sum_{a,x} \Tr(- \tilde{F}^{*}_{a|x}(\alpha)\sigma_{a|x})\\     
    =& \max_{\{\sigma_{a|x}\}\in\mathcal{L}_{\rho}}\frac{\epsilon}{n}\sum_{a,x} \Tr( \tilde{F}^{*}_{a|x}(\alpha)\sigma_{a|x})-\frac{\epsilon}{n}\sum_{a,x} \Tr(\tilde{F}^{*}_{a|x}(\alpha)\tilde{\gamma}^{(\epsilon)}_{a|x}(\alpha))\label{Eq_291}\\ 
    =& \frac{\epsilon}{n}\sum_{a,x}\left[\Tr\left(\tilde{F}^{*}_{a|x}(\alpha)\tilde\sigma^{*}_{a|x}(\alpha)\right)-\Tr( \tilde{F}^{*}_{a|x}(\alpha)\tilde{\gamma}^{(\epsilon)}_{a|x}(\alpha))\right], \label{Eq_301}\\
    =& \frac{\epsilon}{n}\left[\sum_{a,x}\Tr\left({F}^{*}_{a|x}\tilde\sigma^{*}_{a|x}(\alpha)\right)+\alpha \sum_{a,x}\Tr\left(\left(\ketbra{\phi}-\frac{\mathbb{I}}{d}\right)\tilde{\sigma}^{*}_{a|x}(\alpha)\right)-\sum_{a,x}\Tr \left(\left[ F^{*}_{a|x}+\alpha \left(\ketbra{\phi}-\frac{\mathbb{I}}{d}\right)\right] \tilde{\gamma}_{a|x}^{(\epsilon)}(\alpha) \right)\right]\label{45now}\\
    \leq& \frac{\epsilon}{n}\left[1+\alpha \sum_{a,x}\Tr\left(\left(\ketbra{\phi}-\frac{\mathbb{I}}{d}\right)\tilde{\sigma}^{*}_{a|x}(\alpha)\right)-\sum_{a,x}\Tr \left(\left[ F^{*}_{a|x}+\alpha \left(\ketbra{\phi}-\frac{\mathbb{I}}{d}\right)\right] \tilde{\gamma}_{a|x}^{(\epsilon)}(\alpha) \right)\right]\label{45now1}.
\end{align}
Here, we used the identity from Eq.~\eqref{identity} to write Eq.~\eqref{Eq_291} and the optimality of $\tilde\sigma^{*}_{a|x}(\alpha)$ from Eq.~\eqref{tildesigmastar} to arrive at Eq.~\eqref{Eq_301}. Substituting the expression of $\{\tilde{F}^{*}_{a|x}(\alpha)\}_{a,x}$ from Eq.~\eqref{Ftilde} into Eq.~\eqref{Eq_301} yields Eq.~\eqref{45now}, and using the fact that $\tilde{\sigma}^{*}_{a|x}(\alpha)$ is an unsteerable assemblage in $\mathcal{L}_{\rho}$ (see Eq.~\eqref{tildesigmastar}), together with Eq.~\eqref{Lrhomax} and Eq.~\eqref{Eq_unsteerable_witness}, we obtain
\begin{equation}
    \sum_{a,x}\mathrm{Tr}\!\left(F^{*}_{a|x}\,\tilde{\sigma}^{*}_{a|x}(\alpha)\right) \leq 1,
\end{equation}
which in turn implies the inequality in Eq.~\eqref{45now1}. From the expression of $Q_c(\{\rho_{a|x}\},\{H^{(\epsilon)}_{a|x}(\alpha)\})$ obtained in Eq. \eqref{ineq_in_l41} and upper bound of $\max_{\{\sigma_{a|x}\}\in\mathcal{L}_{\rho}}Q_c(\{\sigma_{a|x}\},\{\tilde{H}^{(\epsilon)}_{a|x}(\alpha)\})$ obtained in Eq. \eqref{45now1}, we can further lower bound inequality in Eq. \eqref{ximaxboundintermediate} for all $\epsilon>0$ and $\alpha>0$:
\begin{align}
    \xi_{\rm max}\left(\{\rho_{a|x}\}_{a,x}\right) &\geq \frac{Q_c\bigl(\{\rho_{a|x}\}, \{\tilde{H}^{(\epsilon)}_{a|x}(\alpha)\}\bigr)}
         {\max_{\{\sigma_{a|x}\} \in \mathcal{L}_{\rho}} 
         Q_c\bigl(\{\sigma_{a|x}\}, \{\tilde{H}^{(\epsilon)}_{a|x}(\alpha)\}\bigr)},\nonumber\\
         &\geq \frac{  1+\mathcal{R}(\{\rho_{a|x}\}_{a,x})+\alpha \sum_{a,x}\Tr\left(\left(\ketbra{\phi}-\frac{\mathbb{I}}{d}\right)\rho_{a|x}\right)-\sum_{a,x}\Tr \left(\left[ F^{*}_{a|x}+\alpha \left(\ketbra{\phi}-\frac{\mathbb{I}}{d}\right)\right] \tilde{\gamma}_{a|x}^{(\epsilon)}(\alpha) \right)}{1+\alpha \sum_{a,x}\Tr\left(\left(\ketbra{\phi}-\frac{\mathbb{I}}{d}\right)\tilde{\sigma}^{*}_{a|x}(\alpha)\right)-\sum_{a,x}\Tr \left(\left[ F^{*}_{a|x}+\alpha \left(\ketbra{\phi}-\frac{\mathbb{I}}{d}\right)\right] \tilde{\gamma}_{a|x}^{(\epsilon)}(\alpha) \right)}:=h(\epsilon,\alpha)\label{healpha}.
\end{align}
Observe that for $\epsilon=0$ and $\alpha=0$ we have 
\begin{align}\label{inequality_last_lane567}
    h(0,0)&=\frac{  1+\mathcal{R}(\{\rho_{a|x}\}_{a,x})-\sum_{a,x}\Tr \left(F^{*}_{a|x}p(a|x)\frac{\mathbb{I}}{d} \right)}{1-\sum_{a,x}\Tr \left(F^{*}_{a|x}p(a|x)\frac{\mathbb{I}}{d} \right)}> 1+\mathcal{R}(\{\rho_{a|x}\}_{a,x}),
\end{align}
where to write the final strict inequality in Eq.~\eqref{inequality_last_lane567}, we use the fact that for a steerable assemblage $\{\rho_{a|x}\}_{a,x}$, we have $\mathcal{R}\bigl(\{\rho_{a|x}\}_{a,x}\bigr)>0$ and
\begin{equation}
    \frac{x-z}{y-z}> \frac{x}{y}
    \quad \Longleftrightarrow \quad 
    y-z \geq 0 
    \quad \text{and} \quad 
    x > y,
\end{equation}
with the identifications $x = 1 + \mathcal{R}\bigl(\{\rho_{a|x}\}_{a,x}\bigr)$, $y = 1$ and
$z = \sum_{a,x}p(a|x)\Tr \left( F^{*}_{a|x} \frac{\mathbb{I}}{d}\right)$. Since the inequality is strict at $(\epsilon,\alpha)=(0,0)$, by continuity of the function $h(\epsilon,\alpha)$, we can find a ball of radius $\delta_1>0$ around $(\epsilon,\alpha)=(0,0)$ denoted by $B_{\delta_1}(0,0)$ such that 
\begin{align}\label{eq601}
    \forall (\epsilon,\alpha)\in B_{\delta_1}(0,0)\quad\text{we have}\quad h(\epsilon,\alpha)&=\frac{  1+\mathcal{R}(\{\rho_{a|x}\}_{a,x})+\alpha \sum_{a,x}\Tr\left(\left(\ketbra{\phi}-\frac{\mathbb{I}}{d}\right)\rho_{a|x}\right)-\sum_{a,x}\Tr \left(\left[ F^{*}_{a|x}+\alpha \left(\ketbra{\phi}-\frac{\mathbb{I}}{d}\right)\right] \tilde{\gamma}_{a|x}^{(\epsilon)}(\alpha) \right)}{1+\alpha \sum_{a,x}\Tr\left(\left(\ketbra{\phi}-\frac{\mathbb{I}}{d}\right)\tilde{\sigma}^{*}_{a|x}(\alpha)\right)-\sum_{a,x}\Tr \left(\left[ F^{*}_{a|x}+\alpha \left(\ketbra{\phi}-\frac{\mathbb{I}}{d}\right)\right] \tilde{\gamma}_{a|x}^{(\epsilon)}(\alpha) \right)},\nonumber\\&>1+\mathcal{R}(\{\rho_{a|x}\}_{a,x}).
\end{align}

We now show that, for suitably chosen values $\epsilon > 0$ and $\alpha > 0$, say $(\epsilon,\alpha)=(\tilde{\epsilon}_{*},\tilde{\alpha}_{*})$, the Hamiltonian tuple $\{H^{(\tilde{\epsilon}_*)}_{a|x}(\tilde{\alpha}_{*})\}_{a,x}=\left\{-\tilde{\epsilon}_*\left[F^{*}_{a|x}+\tilde\alpha_{*}\left(\ketbra{\phi}-\frac{\mathbb{I}}{d}\right)\right]\right\}_{a,x}$ defined in Eq.~\eqref{tildeHamiltonian} belongs to $\mathbf{H}^{(\rho)}_{+}$, implying that the heat withdrawn from the bath using this Hamiltonian tuple is non-negative, namely
\begin{equation}\label{comb1}
    Q_c\bigl(\{\rho_{a|x}\},\{H^{(\tilde{\epsilon}_{*})}_{a|x}(\tilde{\alpha}_*)\}\bigr) \geq 0 \quad;\quad 
    \max_{\{\sigma_{a|x}\} \in \mathcal{L}_{\rho}} Q_c\bigl(\{\sigma_{a|x}\}, \{H^{(\tilde{\epsilon}_*)}_{a|x}(\tilde{\alpha}_*)\}\bigr) \geq 0,
\end{equation}
as well as the inequality in Eq.~\eqref{eq601} is also satisfied, i.e., $h(\tilde{\epsilon}_*,\tilde{\alpha}_*) > 1+\mathcal{R}(\{\rho_{a|x}\}_{a,x})$.
First, we will show that $ Q_c(\{\rho_{a|x}\},\{H^{(\epsilon)}_{a|x}(\alpha)\}) \geq 0$ for a suitable values for the parameter $(\epsilon,\alpha)$. We have obtained 
\begin{equation}\label{ineq_in_l412}
    Q_c(\{\rho_{a|x}\},\{H^{(\epsilon)}_{a|x}(\alpha)\})=\frac{\epsilon}{n}f(\epsilon,\alpha),
\end{equation}
in Eq. \eqref{ineq_in_l41}. Evaluating $f(\epsilon,\alpha)$ defined in Eq. \eqref{ineq_in_l41} at $\epsilon = 0$ and $\alpha = 0$ gives
\begin{align}\label{ineq_strict1}
    f(0,0) 
    = 1 + \mathcal{R}(\{\rho_{a|x}\}_{a,x}) - \sum_{a,x} p(a|x)\,\mathrm{Tr}\!\left(F^{*}_{a|x}\frac{\mathbb{I}}{d}\right) 
    = 1 + \mathcal{R}(\{\rho_{a|x}\}_{a,x}) - c \; > \; 0,
\end{align}
where the strict inequality follows from the fact that $c \leq 1$ (see Eq.~\eqref{cineq1}) and $\mathcal{R}(\{\rho_{a|x}\}_{a,x}) > 0$, since the assemblage $\{\rho_{a|x}\}_{a,x}$ is steerable.  Since the inequality in Eq.~\eqref{ineq_strict1} is strict, by continuity of the function \(f(\epsilon,\alpha)\), there exists a ball of radius  \(\delta_2 > 0\) such that, for all \((\alpha,\epsilon)\) in the ball $B_{\delta_2}(0,0)$ where the inequality remains strict i.e.,
\begin{align}\label{ball1}
    \forall(\epsilon,\alpha)&\in B_{\delta_2}(0,0)\quad\text{we have}\quad\nonumber\\f(\epsilon,\alpha)&= 1+\mathcal{R}(\{\rho_{a|x}\}_{a,x})+\alpha \sum_{a,x}\Tr\left(\left(\ketbra{\phi}-\frac{\mathbb{I}}{d}\right)\rho_{a|x}\right)-\sum_{a,x}\Tr \left(\left[ F^{*}_{a|x}+\alpha \left(\ketbra{\phi}-\frac{\mathbb{I}}{d}\right)\right] \tilde{\gamma}_{a|x}^{(\epsilon)}(\alpha) \right)>0.
\end{align}
Eq. \eqref{ball1} allows us to write from Eq. \eqref{ineq_in_l412}:
\begin{align}\label{ball91}
    \forall(\epsilon,\alpha)&\in B_{\delta_2}(0,0)\quad\text{we have}\quad Q_c(\{\rho_{a|x}\},\{H^{(\epsilon)}_{a|x}(\alpha)\}) >0.
\end{align}
Next, we will show that $\max_{\{\sigma_{a|x}\}\in\mathcal{L}_{\rho}}Q_c(\{\sigma_{a|x}\},\{\tilde{H}^{(\epsilon)}_{a|x}(\alpha)\})>0$ for suitably chosen $\epsilon > 0$ and $\alpha > 0$. Consider a sub-optimal assemblage $\{p(a|x)\ketbra{\phi}\}_{a,x}$ where $\ket{\phi}$ is the same normalized state appeared in the definition of Hamiltonian tuple given in Eq. \eqref{tildeHamiltonian} or Eq. \eqref{Ftilde}. From proposition \ref{prop_steerable_assemblage}, we can say $\{p(a|x)\ketbra{\phi}\}_{a,x}\in\mathcal{L}_{\rho}$. Then, from Eq. \eqref{Eq_301} we can have
\begin{align}
    &\max_{\{\sigma_{a|x}\}\in\mathcal{L}_{\rho}}Q_c(\{\sigma_{a|x}\},\{\tilde{H}^{(\epsilon)}_{a|x}(\alpha)\})= \frac{\epsilon}{n}\sum_{a,x}\left[\Tr\left(\tilde{F}^{*}_{a|x}(\alpha)\tilde\sigma^{*}_{a|x}(\alpha)\right)-\Tr( \tilde{F}^{*}_{a|x}(\alpha)\tilde{\gamma}^{(\epsilon)}_{a|x}(\alpha))\right], \label{Eq_302}\\&\geq \frac{\epsilon}{n}\underbrace{\sum_{a,x}\left[\Tr\left(\tilde{F}^{*}_{a|x}(\alpha)p(a|x)\ketbra{\phi}\right)-\Tr( \tilde{F}^{*}_{a|x}(\alpha)\tilde{\gamma}^{(\epsilon)}_{a|x}(\alpha))\right]}_{:=g(\epsilon,\alpha)}\label{Qcstererabletilde23456}= \frac{\epsilon}{n} g(\epsilon,\alpha),
\end{align}
where we write inequality we use Eq. \eqref{tildesigmastar}. For $\epsilon=0$ and for any $\alpha>0$, we have write from Eq. \eqref{Qcstererabletilde23456} 
\begin{align}\label{Le245}
    g(0,\alpha)=&  \sum_{a,x} p(a|x)\Bigg[  \Tr \bigl( F^{*}_{a|x}\ketbra{\phi} \bigr)+\alpha \left(\ketbra{\phi}-\frac{\mathbb{I}}{d}\right)\ketbra{\phi}\Bigg]- \sum_{a,x}p(a|x)\Bigg[  \Tr \left( F^{*}_{a|x}\frac{\mathbb{I}}{d} \right)+\alpha \left(\ketbra{\phi}-\frac{\mathbb{I}}{d}\right)\frac{\mathbb{I}}{d}\Bigg]\nonumber\\
    =&\alpha\left(1-\frac{1}{d}\right)>0,
\end{align}
where to write the final equality we employ the defining condition of case 2 as given in Eq. \eqref{case2}. Since the inequality in Eq.~\eqref{Le245} is strict for any $\alpha>0$, we can choose a $\min\{\delta_1,\delta_2\}>\alpha=\alpha_{+}>0$  where the inequality in Eq. \eqref{Le245} holds strictly. Recall that $\delta_1$ and $\delta_2$ are the radius of ball given in Eq.~\eqref{eq601} and Eq.~\eqref{ball1}. By continuity of the function $g(\epsilon,\alpha)$, there exists a ball of radius $\delta_3>0$ around $(\epsilon,\alpha)=(0,\alpha_{+})$, within which the inequality remains strictly satisfied. Let us denote this ball as $B_{\delta_3}(0,\alpha_{+})$. Thus,
\begin{align}\label{ball234}
    \forall (\epsilon,\alpha) \in B_{\delta_3}(0,\alpha_+) \quad \text{we have}\quad g(\epsilon,\alpha)=\sum_{a,x}\left[\Tr\left(\tilde{F}^{*}_{a|x}(\alpha)p(a|x)\ketbra{\phi}\right)-\Tr( \tilde{F}^{*}_{a|x}(\alpha)\tilde{\gamma}^{(\epsilon)}_{a|x}(\alpha))\right]>0.
\end{align}
Using Eq. \eqref{ball234}, we can conclude that lower bound of $\max_{\{\sigma_{a|x}\}\in\mathcal{L}_{\rho}}Q_c(\{\sigma_{a|x}\},\{\tilde{H}^{(\epsilon)}_{a|x}(\alpha)\})$ obtained in Eq. \eqref{Qcstererabletilde23456} is non-negative, namely
\begin{align}\label{ball2345}
    \forall (\epsilon,\alpha) \in B_{\delta_3}(0,\alpha_+) \quad \text{we have}\quad \max_{\{\sigma_{a|x}\}\in\mathcal{L}_{\rho}}Q_c(\{\sigma_{a|x}\},\{\tilde{H}^{(\epsilon)}_{a|x}(\alpha)\})>0.
\end{align}

To construct a Hamiltonian tuple, we must choose $\epsilon$ and $\alpha$ appropriately so that heat is withdrawn from the bath. To suitably choose  $\epsilon=\tilde\epsilon_{*}$ and $\alpha=\tilde\alpha_{*}$, we simply can take any point with $\epsilon>0$ and $\alpha>0$ from the set 
\begin{equation}
    B_{\delta_1}(0,0)\cap B_{\delta_2}(0,0) \cap  B_{\delta_3}(0,\alpha_+),
\end{equation}
which allows to satisfy Eq. \eqref{eq601}, Eq. \eqref{ball91} and Eq. \eqref{ball2345} simultaneously. We can guarantee  $ B_{\delta_1}(0,0)\cap B_{\delta_2}(0,0) \cap  B_{\delta_3}(0,\alpha_+)$ is non-empty as we have chosen $\alpha_{+}<\min\{\delta_1,\delta_3\}$.  This means the Hamiltonian tuple $\{\tilde{H}_{a|x}^{*}\}_{a,x}=\{H^{(\tilde\epsilon_{*})}_{a|x}(\tilde\alpha_{*})\}_{a,x}\in \textbf{H}^{(\rho)}_{+}$ (follows from Eq. \eqref{ball91} and Eq. \eqref{ball2345}) and combining with Eq. \eqref{healpha} and Eq. \eqref{eq601} we can say that
\begin{equation}\label{rob_ineq}
    \xi_{\rm max}\left(\{\rho_{a|x}\}_{a,x}\right)\geq h(\tilde\epsilon_{*},\tilde\alpha_{*})>1+\mathcal{R}(\{\rho_{a|x}\}_{a,x}).
\end{equation}
This completes the proof for case 2.
\cblue

\cblue
\section{Temperature dependence of the advantages in cooling}
Theorem 1 establishes that, for any finite and nonzero inverse temperature $\beta$ of the bath and any given steerable assemblage $\{\rho_{a|x}\}_{a,x}$ used as a resource, one can always construct a Hamiltonian tuple that enables a greater heat withdrawal fueled by the steerable assemblage  $\{\rho_{a|x}\}_{a,x}$ compared to all unsteerable assemblages in the set $\mathcal{L}_{\rho}$. For instance, the parameter $\epsilon_*$ in the Hamiltonian tuple in Eq.~\eqref{Hamiltonian_from_witness} considered in case 1, or the parameters $(\tilde{\epsilon}_{*},\tilde{\alpha}_{*})$ in Eq.~\eqref{comb1} considered in case 2, depend on the bath temperature, since they are chosen such that heat can be withdrawn from the bath, thereby ensuring that the inequality in Eq.~\eqref{stareqn} (for case 1) or Eq.~\eqref{comb1} (for case 2) is satisfied. In this section, we are going to analyze a tighter bound on the cooling advantage compared to the one presented in Theorem 1 of the main text.

\begin{prop}
Consider a cooling task fueled by a steerable assemblage $\{\rho_{a|x}\}_{a,x}$ and the Hamiltonian tuple
\begin{equation}
    \{H_{a|x}^{*}\}_{a,x}=\{-\epsilon_*F^*_{a|x}\}_{a,x}
    \quad \text{with} \quad 
    \sum_{a,x} p(a|x)F^{*}_{a|x} \not\propto \mathbb{I},
\end{equation}
such that
\begin{align}\label{stareqnre}
\max_{\{\sigma_{a|x}\} \in \mathcal{L}_{\rho}}Q_c\bigl(\{\sigma_{a|x}\}, \{H^{*}_{a|x}\}\bigr) \geq 0
\quad \text{and} \quad 
Q_c\bigl(\{\rho_{a|x}\}, \{H^{*}_{a|x}\}\bigr) \geq 0,
\end{align}
for a range of inverse bath temperatures
\begin{equation}\label{temperature_range}
    \beta_{\min}\leq \beta\leq \beta_{\max}.
\end{equation}
That is, within the temperature range specified in Eq.~\eqref{temperature_range}, the Hamiltonian $\{H^{*}_{a|x}\}_{a,x}$ together with the steerable assemblage $\{\rho_{a|x}\}_{a,x}$ enables heat withdrawal from the bath. Then, the lower bound on $\xi_{\max}(\{\rho_{a|x}\}_{a,x})$ from Eq.~\eqref{last_step_app}, namely
\begin{align}
    \xi_{\rm max}(\{\rho_{a|x}\}_{a,x})
    \geq
    \frac{ 1+\mathcal{R}(\{\rho_{a|x}\}_{a,x}) - \sum_{a,x}\Tr \bigl( F^{*}_{a|x} \gamma^{(\epsilon_*)}_{a|x} \bigr) }
    { 1 - \sum_{a,x} \Tr \bigl( F^{*}_{a|x} \gamma^{(\epsilon_*)}_{a|x}\bigr)}:=B^{(\epsilon_*)}(\beta) 
    \label{inequality_last_lane2}
\end{align}
is an increasing function of the inverse temperature $\beta$. This implies that the quantum advantage in cooling is expected to be more pronounced at lower temperatures (i.e., larger $\beta$) compared to higher temperatures.
\end{prop}
\begin{proof}
    We proceed by observing that the right-hand side of the inequality given in Eq.~\eqref{inequality_last_lane2} depends on the temperature, since $\{\gamma^{(\epsilon_*)}_{a|x}\}_{a,x}$ is the collection of sub-normalized  Gibbs state at the inverse temperature of the bath, $\beta$, with Hamiltonian tuple $\{H_{a|x}\}_{a,x} = \{-\epsilon_* F^{*}_{a|x}\}_{a,x}$ i.e.,
\begin{align}
    \gamma^{(\epsilon_*)}_{a|x}&:=p(a|x)\hat\gamma^{(\epsilon_*)}_{a|x},\quad\text{where}\quad p(a|x)=\Tr(\rho_{a|x}) \quad\text{and} \label{Eq_Gibbs_Assemblages}\\ \;\; \hat\gamma^{(\epsilon_*)}_{a|x}&:=\frac{e^{-\beta H^{*}_{a|x}}}{\Tr\left(e^{-\beta H^{*}_{a|x}}\right)}=\frac{e^{\epsilon_*\beta  F^{*}_{a|x}}}{\Tr\left(e^{\epsilon_* \beta F^{*}_{a|x}}\right)}\label{Eq_Gibbs_Assemblages_state}.
\end{align}
 Now, we will show that the right hand side of the inequality in Eq. \eqref{inequality_last_lane2} is an increasing function of inverse temperature $\beta$. In order to show this, we proceed by showing the derivative with respect to inverse temperature $\beta$ is non-negative as follows:
\begin{align}
    &\frac{d}{d\beta}\left(\frac{ \left[ 1+\mathcal{R}(\{\rho_{a|x}\}_{a,x}) - \sum_{a,x}\Tr \bigl( F^{*}_{a|x} \gamma^{(\epsilon_*)}_{a|x} \bigr) \right]}{\left[1 - \sum_{a,x} \Tr \bigl( F^{*}_{a|x} \gamma^{(\epsilon_*)}_{a|x}\bigr) \right]}\right)\nonumber\\=& \frac{\mathcal{R}(\{\rho_{a|x}\}_{a,x})}{\left[1 - \sum_{a,x} \Tr \bigl( F^{*}_{a|x} \gamma^{(\epsilon_*)}_{a|x}\bigr) \right]^2}\times \left(\sum_{a,x}\frac{d}{d\beta}\Tr \bigl( F^{*}_{a|x} \gamma^{(\epsilon_*)}_{a|x} \bigr)\right),\nonumber\\
=& \frac{\epsilon_*\mathcal{R}(\{\rho_{a|x}\}_{a,x})}{\left[1 - \sum_{a,x} \Tr \bigl( F^{\epsilon_*}_{a|x} \gamma^{(\epsilon_*)}_{a|x}\bigr) \right]^2}\times \sum_{a,x} p(a|x) \text{Var}_{\hat\gamma^{(\epsilon_*)}_{a|x}}(F^{*}_{a|x})\geq 0\label{inequality_inv_temp},
\end{align}
where $\text{Var}_{\hat\gamma^{(\epsilon_*)}_{a|x}}(F^{*}_{a|x})$ is the variance of the operator $F^{*}_{a|x}$ with respect to Gibbs state $\hat\gamma^{(\epsilon_*)}_{a|x}$ defined Eq. \eqref{Eq_Gibbs_Assemblages_state} which is always non-negative  i.e.,
\begin{equation}
   \forall A,\hat{\rho},\quad\text{Var}_{\hat\varrho}(A):=\Tr(A^2\hat\varrho)-(\Tr(A\hat\varrho))^2\geq 0,
\end{equation}
where $\hat{\varrho}$ is the normalized quantum state. Thus, from the inequality in Eq.~\eqref{inequality_inv_temp}, we observe that the bound on the cooling advantage in Eq.~\eqref{inequality_last_lane2} increases with the inverse temperature, which means
\begin{equation}
    B^{(\epsilon)}(\beta_{\min})\leq\frac{ 1+\mathcal{R}(\{\rho_{a|x}\}_{a,x}) - \sum_{a,x}\Tr \bigl( F^{*}_{a|x} \gamma^{(\epsilon_*)}_{a|x} \bigr) }
    { 1 - \sum_{a,x} \Tr \bigl( F^{*}_{a|x} \gamma^{(\epsilon_*)}_{a|x}\bigr)}:=B^{(\epsilon_*)}(\beta)\leq B^{(\epsilon)}(\beta_{\max}).
\end{equation}
Equivalently, this implies that the bound in Eq.~\eqref{inequality_inv_temp} decreases as the temperature increases. In other words, at low bath temperatures, the amount of heat that can be withdrawn using unsteerable assemblages as fuel can be significantly smaller compared to that achievable with steerable assemblages. 

\end{proof}

A similar behavior can also be proven in the case where $\sum_{a,x} p(a|x)\, F^{*}_{a|x} \propto \mathbb{I}$, and can be established straightforwardly by partially differentiating $h(\tilde{\epsilon}_*,\tilde{\alpha}_*)$ in Eq.~\eqref{healpha} with respect to $\beta$ and showing that the derivative is non-negative. For the nuclear spin in the nitrogen-vacancy (NV) center considered in the end matter, we observe the same qualitative behavior: the lower bound of $\xi_{\rm max}(\{\rho_{a|x}\}_{a,x})$ increases with inverse temperature. For the corresponding energy scale $\approx (0.1 \;\text{MHz}-10 \;\text{MHz})$, the inverse temperature is bounded as
\begin{equation}\label{beta_bound_supp}
    0 \leq \beta \leq \frac{\ln 2}{2\epsilon} = 8.37 \times 10^7\ \text{eV}^{-1}.
\end{equation}

\blk

\section{Satisfiability of the constraint under the chosen suboptimal operators } Here, we prove that
\begin{equation}\label{Fax_example}
    F_{a|x}=\frac{1}{1+\sqrt{d}}\ketbra{\phi^a_x}{\phi^a_x}\quad\quad \forall a, x,
\end{equation}
satisfies the constraint of the SDP given in Eq.~\eqref{Dual_robustness},  where $a\in\{0,\ldots,d-1\}$ and $x\in\{0,\ldots,d\}$. 
The operators \(F_{a|x}\) defined in Eq.~\eqref{Fax_example} are positive semidefinite for all \(a\) and \(x\). Next, we show that
$\sum_{a,x}D(a|x,\lambda) F_{a|x}\leq \mathbb{I}$ for all $\lambda$. This is equivalent to showing that
\begin{equation}\label{ineq_Operator_norm}
     \left\|\sum_{a,x}D(a|x,\lambda) F_{a|x}\right\|_{\infty}\leq 1\qquad \forall \lambda,
\end{equation}
where \(D(a|x,\lambda)=\delta_{a,\lambda(x)}\) denotes the deterministic response function.  

Before proving Eq.~\eqref{ineq_Operator_norm}, we establish a relation that will be used in the proof. Consider a bipartite pure state on systems \(C\) and \(D\) of the form 
\begin{equation}
    \ket{\zeta}_{CD} = \sum_{x=1}^{d+1} \ket{\phi^{\lambda(x)}_{x}}_C\ket{x}_D,
\end{equation}
where the $\ket{x}$’s form an orthonormal basis. Using the Schmidt decomposition of \(\ket{\zeta}_{CD}\), we can infer that 
\begin{align}
    &\rm Spec\left[\Tr_{C}\left(\ket{\zeta}_{CD}\right)\right] = \rm Spec\left[\Tr_{D}\left(\ket{\zeta}_{CD}\right)\right]\nonumber\\
    \Longleftrightarrow \;& \rm Spec\left[\sum_{x}\ketbra{\phi^{\lambda(x)}_x}{\phi^{\lambda(x)}_x}\right] = \rm Spec\left[\sum_{x,y}\braket{\phi^{\lambda(y)}_y}{\phi^{\lambda(x)}_x}\ketbra{y}{x}\right]\nonumber\\
    \Longleftrightarrow \;&\left\|\sum_{x}\ketbra{\phi^{\lambda(x)}_x}{\phi^{\lambda(x)}_x}\right\|_{\infty}= \left\|\sum_{x,y}\braket{\phi^{\lambda(y)}_y}{\phi^{\lambda(x)}_x}\ketbra{y}{x}\right\|_{\infty},\label{Eq:equality_of_ineq}
\end{align}
where \(\rm Spec[\cdot]\) denotes the spectrum and \(\|\cdot\|_{\infty}\) the operator norm.  Then, the proof of Eq.~\eqref{ineq_Operator_norm} follows:
\begin{align}
      &\left\|\sum_{a,x}D(a|x,\lambda) F_{a|x}\right\|_{\infty}=\frac{1}{1+\sqrt{d}}\left\|\sum_{a,x}D(a|x,\lambda) \ketbra{\phi^a_x}{\phi^a_x}\right\|_{\infty}\nonumber\\
      &= \frac{1}{1+\sqrt{d}}\left\|\sum_{a,x}\delta_{a,\lambda(x)} \ketbra{\phi^a_x}{\phi^a_x}\right\|_{\infty}= \frac{1}{1+\sqrt{d}}\left\|\sum_{x}\ketbra{\phi^{\lambda(x)}_x}{\phi^{\lambda(x)}_x}\right\|_{\infty}\nonumber\\
      &= \frac{1}{1+\sqrt{d}}\left\|\sum_{x,y}\braket{\phi^{\lambda(y)}_y}{\phi^{\lambda(x)}_x}\ketbra{y}{x}\right\|_{\infty}\nonumber\\
      &=\frac{1}{1+\sqrt{d}}\left\|\sum_{x}\braket{\phi^{\lambda(x)}_x}{\phi^{\lambda(x)}_x}\ketbra{x}{x}+\sum_{x\neq y}\braket{\phi^{\lambda(y)}_y}{\phi^{\lambda(x)}_x}\ketbra{y}{x}\right\|_{\infty}\nonumber\\ 
      &=\frac{1}{1+\sqrt{d}}\left\|\sum_{x}\ketbra{x}{x}+\sum_{x\neq y}\frac{1}{\sqrt{d}}e^{i\phi_{x,y}}\ketbra{y}{x}\right\|_{\infty}\nonumber\\ 
      &=\frac{1}{1+\sqrt{d}}\left\|\sum_{x}\left(1-\frac{1}{\sqrt{d}}\right)\ketbra{x}{x}+\sum_{x,y}\frac{1}{\sqrt{d}}e^{i\phi_{x,y}}\ketbra{y}{x}\right\|_{\infty}\nonumber\\
      &\leq\frac{1}{1+\sqrt{d}}\left\|\sum_{x}\left(1-\frac{1}{\sqrt{d}}\right)\ketbra{x}{x}\right\|_{\infty}+\left\|\sum_{x,y}\frac{1}{\sqrt{d}}e^{i\phi_{x,y}}\ketbra{y}{x}\right\|_{\infty}\nonumber\\
      &\leq\frac{1}{1+\sqrt{d}}\left\|\sum_{x}\left(1-\frac{1}{\sqrt{d}}\right)\ketbra{x}{x}\right\|_{\infty}+\left\|\sum_{x,y}\frac{1}{\sqrt{d}}e^{i\phi_{x,y}}\ketbra{y}{x}\right\|_{F}\\
      &= \frac{1}{1+\sqrt{d}}\left[\left(1-\frac{1}{\sqrt{d}}\right)+\frac{1}{\sqrt{d}}(d+1)\right]=1,
\end{align}
where the fifth equality follows from Eq.~\eqref{Eq:equality_of_ineq}, and in the sixth inequality we use the property of mutually unbiased bases (MUBs),
\begin{equation}
    \braket{\phi^{\lambda(y)}_y}{\phi^{\lambda(x)}_x}=\frac{1}{\sqrt{d}}e^{i\phi_{x,y}} \qquad \phi_{x,y}\in \mathbb{R}.
\end{equation}
\cblue We write the first inequality using the triangle inequality \blk while the final inequality follows from the relation \(\|\cdot\|_{\infty}\leq \|\cdot\|_{F}\), where the Frobenius norm is defined as
\begin{equation}
\|A\|_F = \sqrt{\operatorname{Tr}(A^\dagger A)} = \sqrt{\sum_{i=1}^{m}\sum_{j=1}^{n} |A_{ij}|^2}.
\end{equation}

\section{Lower bound on the steerability robustness for an assemblage generated from an isotropic state}\label{app:Isotropic}
In this section, we evaluate the robustness of the assemblages obtained from the isotropic state by performing measurements  on one subsystem (See Eq. (17) of the main text):
\begin{equation}\label{MUB_measurement_assemblage_2}
    \left\{\ketbra{\phi^{*a}_{x}}{\phi^{*a}_{x}}\right\}_{a=0}^{d-1},
    \qquad 
    \ket{\phi^{*a}_{x}} = \sum_{i=0}^{d-1} \braket{\phi^a_x}{i}\ket{i}.
\end{equation}
 Here, $\ket{\phi^a_x}$ denotes the $a$-th vector of the mutually unbiased basis (MUB) indexed by $x$, satisfying
\begin{equation}
\bigl|\braket{\phi^a_x}{\phi^b_y}\bigr| =
\begin{cases}
\delta_{ab}, & x = y,\\[2mm]
1/\sqrt{d}, & x \neq y.
\end{cases}
\end{equation} 
The isotropic state is defined as a convex mixture of a maximally entangled state and the maximally mixed state. Explicitly, it can be written as
\begin{equation}\label{Eq:Isotropic}
\rho^{\mathrm{isotropic}}_{AB}(\eta)
=\eta\left(\frac{1}{d}\sum_{i=0}^{d-1}\sum_{j=0}^{d-1} 
\ketbra{i}{j}_{A}\otimes\ketbra{i}{j}_{B}\right)
+(1-\eta)\frac{\mathbb{I}_{AB}}{d^2},
\end{equation}
where the parameter \(\eta \in [0,1]\) quantifies the weight of the maximally entangled component in the mixture, while the remaining contribution corresponds to the maximally mixed state. We see from Eq.~\eqref{Eq:Isotropic} that when $\eta = 1$, 
$\rho_{\mathrm{isotropic}}(\eta)$ reduces to the maximally entangled state, 
and when $\eta = 0$, $\rho_{\mathrm{isotropic}}(\eta)$ reduces to the maximally mixed state. We can calculate the assemblage obtained by measuring subsystem $A$ 
in the state $\ket{\phi^{a}_{x}}$, which yields for all $a$ and $x$:
\begin{eqnarray}\label{isotropic_assemblage}
    \Tr_A\!\left[\left(\ketbra{\phi^{*a}_{x}}{\phi^{*a}_{x}}\right)
    \rho^{\mathrm{isotropic}}_{AB}(\eta)\right]
    &=& \eta\, \frac{1}{d}\ketbra{\phi^{a}_{x}}_{B}
    + (1 - \eta)\,\frac{\mathbb{I}_{B}}{d^2}\nonumber\\
    &=& \eta\, \rho^{\rm max.\; ent.}_{a|x}
    + (1 - \eta)\,\frac{\mathbb{I}}{d^2}\nonumber\\
    &:=& \rho^{\rm isotropic}_{a|x} (\eta)
\end{eqnarray}
where we have dropped the subscript for brevity to write the second equality. In Refs.~\cite{Wiseman_steering, SteeringRevModPhys}, it has been shown that when 
\begin{equation}
    \eta \leq \frac{H_d - 1}{d - 1},
\end{equation}
the assemblage $\rho^{\rm isotropic}_{a|x} (\eta)$ given in Eq.~\eqref{isotropic_assemblage} is unsteerable, whereas it becomes steerable for 
\begin{equation}
    \eta > \frac{H_d - 1}{d - 1},
\end{equation}
where $H_d \coloneqq \sum_{i=1}^d i^{-1}$ denotes the $d$-th harmonic number. In the following, we calculate a lower bound on the steerability robustness for the assemblage $\rho^{\rm isotropic}_{a|x} (\eta)$. 

To lower bound the steerability robustness, we employ the SDP given in Eq.~\eqref{Dual_robustness}. We consider the same sub-optimal choice for $F_{a|x}$ as in Eq.~\eqref{Fax_example}.
Substituting such a suboptimal choice for $F_{a|x}$ from Eq.~\eqref{Fax_example}, we can calculate the lower bound of steerability robustness as follows:
\begin{align}
    \mathcal{R}\left(\left\{\rho^{\rm isotropic}_{a|x}(\eta)\right\}_{a,x}\right)  &\geq\sum_{a,x}\Tr\left(F_{a|x}\rho^{\rm isotropic}_{a|x}(\eta)\right)-1\nonumber\\
    &= \sqrt{d}\left(\frac{\sqrt{d}-1}{\sqrt{d}+1}\right)\eta-\left(\frac{d\sqrt{d}-1}{d\left(\sqrt{d}+1\right)}\right)(1-\eta).\label{Lower_bound_robustness_isotropic}
\end{align}
The prefactor of $\eta$ in Eq.~\eqref{Lower_bound_robustness_isotropic} scales as
\begin{equation}
    \sqrt{d} \left( \frac{\sqrt{d}-1}{\sqrt{d}+1} \right) = \mathcal{O}(\sqrt{d}),
\end{equation}
while the prefactor of $(1-\eta)$, satisfies
\begin{equation}
    \frac{d\sqrt{d}-1}{d\left(\sqrt{d}+1\right)} < 1 = \mathcal{O}(1).
\end{equation}

The quantum advantage in cooling is observed when the lower bound in Eq.~\eqref{Lower_bound_robustness_isotropic} is strictly positive (See Eq. 11 of the main text), i.e.,
\begin{align}
    & \sqrt{d} \left( \frac{\sqrt{d}-1}{\sqrt{d}+1} \right)\eta -  \left( \frac{d \sqrt{d}-1}{d(\sqrt{d}+1)} \right)(1-\eta) > 0 
    \\
    &\Longleftrightarrow \quad \eta > \frac{1}{\sqrt{d} + \frac{1}{d+\sqrt{d}+1}}=\mathcal{O}\left(\frac{1}{\sqrt{d}}\right).
    \label{eta_bound_positive}
\end{align}
Hence, for any \(\eta\) satisfying Eq.~\eqref{eta_bound_positive}, the lower bound on the steerability robustness of the assemblage \(\rho^{\mathrm{isotropic}}_{a|x}(\eta)\) in Eq.~\eqref{Lower_bound_robustness_isotropic} remains positive, thereby ensuring a quantum advantage in cooling. Moreover, we observe that this lower bound decreases with increasing system dimension, indicating that a smaller admixture of the maximally entangled state is sufficient to achieve a quantum advantage in higher dimensions.

In particular, for \(\eta = 1\), the isotropic state reduces to the maximally entangled state, yielding
\begin{equation}
\rho^{\mathrm{isotropic}}_{a|x}(\eta=1) = \frac{1}{d} \ketbra{\phi^a_x}{\phi^a_x} = \rho^{\mathrm{max.\,ent.}}_{a|x}, 
\qquad \forall a, x.
\end{equation}
In this case, the lower bound in Eq.~\eqref{Lower_bound_robustness_isotropic} simplifies to
\begin{equation}
\sqrt{d}\left(\frac{\sqrt{d}-1}{\sqrt{d}+1}\right) = \mathcal{O}(\sqrt{d}).
\end{equation}
Consequently, the quantum advantage in cooling, (See Eq. 11 of the main text) scales as \(\mathcal{O}(\sqrt{d})\). Such an enhancement, which grows with the dimension of the underlying system, is sometimes referred to as an ``unbounded quantum advantage".

To establish a general condition for obtaining a dimension-dependent advantage, we seek the range of values of \(\eta\) for which the lower bound is not only positive but also increases with the system dimension. Earlier, in Eq.~\eqref{eta_bound_positive}, we identified the region of \(\eta\) that ensures the positivity of the lower bound in Eq.~\eqref{Lower_bound_robustness_isotropic}. We now determine the range of \(\eta\) for which this lower bound grows with dimension.

To this end, we impose that the first-order derivative of the lower bound with respect to \(d\) be positive:
\begin{equation}
\frac{\partial}{\partial d}\!\left[\sqrt{d}\!\left(\frac{\sqrt{d}-1}{\sqrt{d}+1}\right)\!\eta 
- \!\left(\frac{d\sqrt{d}-1}{d(\sqrt{d}+1)}\right)\!(1-\eta)\right] > 0.
\end{equation}
Solving this inequality yields
\begin{equation}
\eta > \frac{2 + \sqrt{d}(3 + d)}{2 + 3\sqrt{d} + 2d^2 + d^{5/2}} = \mathcal{O}\!\left(\frac{1}{d}\right).
\end{equation}
It is straightforward to verify that
\begin{equation}
\frac{1}{\sqrt{d} + \frac{1}{d + \sqrt{d} + 1}} 
> 
\frac{2 + \sqrt{d}(3 + d)}{2 + 3\sqrt{d} + 2d^2 + d^{5/2}}.
\end{equation}
This implies that whenever \(\eta\) scales as
\begin{equation}
\eta > \frac{1}{\sqrt{d} + \frac{1}{d + \sqrt{d} + 1}},
\end{equation}
one not only achieves a quantum advantage in cooling but also an advantage that scales with the dimension of the underlying system.

\end{document}